\documentclass[12pt, a4paper]{article}
\usepackage[utf8]{inputenc}

\usepackage{booktabs}
\usepackage{afterpage}
\usepackage[T1]{fontenc}
\usepackage{textcomp}
\usepackage{fullpage}
\usepackage{graphicx,mathptmx,amsmath,amsfonts,amscd,amsthm,amssymb,eucal,psfrag,color,subfig,url}
\usepackage{defs}
\usepackage{authblk}

\newcommand{\pt}{\ensuremath{p_{\text T}}\:}
\newcommand{\met}{\ensuremath{E_{\text T}^{\text{miss}}}\:}

\begin{document}

\title{HE-LHC prospects for diboson resonance searches and electroweak $WW/WZ$ production via vector boson scattering in the semi-leptonic final states}
\author[1]{Viviana Cavaliere}

\author[2]{Robert Les}
\author[3]{Tatsumi Nitta}
\author[3]{Koji Terashi}

\affil[1]{Brookhaven National Laboratory}
\affil[2]{University of Toronto}
\affil[3]{University of Tokyo}

\maketitle

\abstract{
This note presents the prospects of searches for new heavy resonances decaying to diboson ($WW$) and measurements of electroweak $WW/WZ$ production via vector boson scattering (VBS) 
in association with a high-mass 
dijet system in the $\ell\nu qq$  final states ($W\to\ell\nu$, $W/Z\to qq$).
The prospects are presented for an integrated luminosity of 15~ab$^{-1}$ of proton-proton ($pp$) collisions at 
$\sqrt{s}=27$~TeV with an ATLAS-like detector simulated in the Delphes framework. 
The cross-section measurement of the electroweak $WW/WZ$ production in VBS processes 
is expected to reach the precision of $\sim$2-3\%, improving the expected accuracy at the HL-LHC by a factor of 2. Prospects are presented also for the separation of the longitudinal component of the electroweak $WW/WZ$ production, showing the expected significance of $\sim3\sigma$ is reached with 3~ab$^{-1}$ for the  single $\ell \nu qq$ channel 
and $\sim5\sigma$ for all the semi-leptonic channels combined.
The diboson resonance searches are interpreted for sensitivity to a simplified phenomenological model 
with a heavy gauge boson.  With 15~ab$^{-1}$ of $pp$ data, the discovery reach for the new resonance is 
extended to 8~TeV. 
}

\newpage
\tableofcontents

\section{Introduction}
%
Since the discovery of a Higgs boson at the Large Hadron Collider (LHC) at CERN, the properties of the discovered particle
have been measured, indirectly probing the existence of physics beyond the Standard Model (SM) through Higgs 
production and decay processes. 
The measured properties of the Higgs boson have been so far in good agreement with SM predictions; the coupling strength 
to SM fermions and bosons are confirmed within the level of 10-20\%.
The spin-0 nature of the Higgs boson casts however a big challenge on particle physics, 
requiring a natural explanation to the stability of the measured Higgs mass under the large hierarchy between 
the electroweak and fundamental gravity scales (so-called hierarchy problem).  
Many scenarios beyond the Standard Model (BSM) to explain the hierarchy problem, 
such as the Randall--Sundrum (RS) model with a warped extra dimension~\cite{RSmodel},
models with extended Higgs sectors in the two-Higgs-doublet model (2HDM)~\cite{Branco:2011iw},
models with composite Higgs bosons~\cite{Contino:2011np} 
or extended gauge sectors in Grand Unified Theories~\cite{Pati:1974yy,Georgi:1974sy,Fritzsch:1974nn}, 
postulate the existence of new heavy particles. 
Search for new massive particles has been an important part of the physics program at the LHC, performed extensively 
over a broad range of final states at the centre-of-mass energies $\sqrt{s}$ of 7, 8 and 13~TeV. However, 
no clear hint of such new particles has been observed yet.

While the presence of resonances is the most dramatic signal for new phenomena, they may be too heavy or broad to be clearly observed. The study
of high-energy scattering between the longitudinal components of the vector bosons (vector boson scattering or VBS) is a perfect probe for the presence of new particles or interactions behind the electroweak (EW) symmetry breaking.  In fact, the scattering amplitude of VBS processes could exhibit anomalous behaviour at high energy, in presence of any small deviation of Higgs boson couplings from SM predictions. Such behaviour could give an additional insight to the presence of new resonances at TeV scale, that may play a role in restoring the SM scattering amplitudes. Experimental test for the high-energy behaviour of VBS processes is 
one of the most important measurements at the HL-LHC.

The previous VBS measurements and searches for anomalous quartic gauge couplings (aQGCs) have focused on channels involving leptonic boson decays ($W\to\ell\nu$ and $Z\to\ell\ell$)~\footnote{Unless otherwise noted, $\ell$ stands for electron or muon in this note.} and photons.
The semi-leptonic channels, i.e, $\Vqq\Zvv$, $\Vqq\Wlv$ and $\Vqq\Zll$ where $V=W$ or $Z$, however, can offer an interesting
advantage over the leptonic channels to being able to probe VBS processes with high-\pt vector bosons due to larger $\Vqq$ branching fractions. The hadronically-decaying $V$ boson can be fully reconstructed and identified using the jet substructure techniques. Exploring the VBS topology at a TeV or higher scale will provide an unique opportunity to directly 
test the SM description of the EW symmetry breaking and the role of Higgs particle.

The quest for new heavy particles and anomalous VBS processes will continue in future runs of the LHC and high-luminosity LHC (HL-LHC). 
However, the improvement on the mass reach for new particles is going to be marginal 
due to available parton-parton energy limited by the maximum $\sqrt{s}$ of 14~TeV 
at the LHC and HL-LHC. 
A proposal to extend the LHC beam energy by almost factor 2, called high-energy LHC (HE-LHC), will significantly
improve sensitivity to new massive particles and VBS processes at multi-TeV scale.

This note presents expected sensitivities, at the integrated luminosity of 15~ab$^{-1}$  of $pp$ collisions at $\sqrt{s}=27$~TeV,
of the search for resonances decaying to diboson ($WW$) and VBS measurements in the semi-leptonic channel,
where one $W$ boson decays leptonically and the other $W$ or $Z$ boson decays to quarks ($\ell\nu qq$ channel). 
Both analyses are based on event selection and classification similar to those used in the Run 2 ATLAS searches for 
$VV$ resonances and VBS at $\sqrt{s}=13$~TeV~\cite{EXOT-2016-10, EXOT-2016-28}. 
The main signal and background processes are reconstructed using a generic detector
in the Delphes simulation framework~\cite{Delphes}, modeled from the ATLAS detector in the HL-LHC.
It is assumed that resonance searches in other semi-leptonic and fully hadronic decay channels have similar sensitivities to the 
$\ell\nu qq$ channel at high masses, as observed in the ATLAS $VV$ resonance search~\cite{EXOT-2016-28}.
ATLAS has presented results of VBS searches in the leptonic final states of the $W^{\pm}W^{\pm}$~\cite{ATLAS-CONF-2018-030} and $WZ$~\cite{ATLAS-CONF-2018-033} channels with the observation 
of $6.9\sigma$ and $5.6\sigma$ significance, respectively. ATLAS has also recently presented prospects for resonance and VBS searches in the $\ell\nu qq$ channel, assuming an integrated luminosity of 300 or 3000~fb$^{-1}$ of $pp$ collisions at $\sqrt{s}=14$~TeV~\cite{HLLHC_PUBnote}.

 
\section{Simulation samples}
%
\subsection{Signal simulation}
\label{sec:mc_signal}

The prospect for resonance searches presented in this article is interpreted in the context of the  heavy vector triplet (HVT) model~\cite{HVT,Pappadopulo:2014qza}.
The HVT model provides a broad phenomenological framework to test a  range of
different scenarios involving new heavy gauge bosons and their couplings to SM fermions and bosons.
In this model, a triplet $\mathcal{W}$ of colorless vector bosons is introduced with zero hypercharge.
This leads to a set of nearly-degenerate charged $W^{\prime \pm}$ and neutral $Z^{\prime}$
states, collectively denoted by $V^\prime$, whose masses are assumed to be the same.
A HVT scenario with Drell-Yan production of $V^\prime$ boson, referred to as model A, 
represents the phenomenology of weakly coupled models based on an extended gauge symmetry~\cite{Barger:1980ix},
and is used as a benchmark for interpreting results.

The HVT signal events are generated with \MGMCatNLO ~v2.6.3~\cite{madgraph}
at leading order (LO) using the NNPDF23LO parton distribution function (PDF) set~\cite{Ball:2012cx}. 
The generated events are interfaced 
to \PYTHIAV{8.23}~\cite{pythia8} for parton showering, hadronization, and the underlying event.

The electroweak \VVjj\ production is modeled with \MGMCatNLO~v2.6.3~\cite{Alwall:2014hca},
interfaced to \PYTHIAV{8.23}~\cite{Sjostrand:2007gs} for parton showering and hadronization.
The  \textsc{NNPDF30LO} PDF set~\cite{Ball:2012cx} is used.
The electroweak \VVjj\ samples are generated with two on-shell $V$ bosons, with one $W$ boson decaying leptonically
($W\to \ell \nu$ with $\ell= e, \mu, \tau$), 
and the other $V$ boson decaying hadronically.
For each sample, all of the purely electroweak tree-level diagrams at $\mathcal{O}(\alpha_{EW}^6)$
that contribute to the final state are included, i.e, VBS diagrams, non-VBS electroweak diagrams with and without $b$-quarks in the initial or final states.
The non-VBS diagrams, e.g, diagrams including a $Wtb$ vertex,  are suppressed by requiring the tagging jets not to
 be $b$-tagged ($b$-veto) in the analysis, as described in Section.~\ref{sec:analysis}.
For electroweak $WWjj$ production, the electroweak $t\bar{t}$ processes have a significant contribution~($\sim$ 70\%).  Such contribution can be effectively removed by applying a $b$-jet veto.
Diagrams that contain a mixture of electroweak and QCD vertices ($\mathcal{O}(\alpha_{EW}^4 \alpha_{S}^2)$ diagrams) are not included in these samples,
and are not part of the signal definition.  Such processes  are accounted for by the background samples of $\ttbar$, single-top, and diboson production.

For the extraction of the longitudinal components in VBS processes, the electroweak $WWjj$ samples are generated with the {\it DECAY} program to identify the polarization state of the produced $V$ bosons. 
The generated events are then classified according to the polarization state: both $V$ bosons are longitudinally (LL) 
or transversely (TT) polarized, or in the mixed state (LT). Each event is showered using  \PYTHIAV{8.23}~\cite{Sjostrand:2007gs}  and 
then processed through the Delphes simulation.

\subsection{Background simulation}
\label{sec:mc_bkg}

The main background in this analysis is $W$ boson produced in association with jets ($W$+jets), with a significant contribution from top-quark pair production.
The $W$+jets events are simulated using \MGMCatNLO ~v2.6.2 at LO using the NNPDF30NLO PDF set, interfaced to \PYTHIAV{8.23} for parton showering and hadronization. 
The $W\to \tau\nu$ events are included in the $W$+jets sample.
For the generation of top-quark pairs, the \textsc{aMC@NLO} event generator is used with the NNPDF30NLO PDF set, interfaced to \PYTHIAV{8.23}for parton showering and hadronization.
The top quark mass is set to 172.5 GeV.  
The $Z$+jets, single-top and non-resonant diboson ($ZZ$, $WZ$ and $WW$) processes 
are not simulated and are expected to contribute at most 10\% to the total background.
 
\section{Event selection}
%
\label{sec:analysis}
Search for diboson resonances and the measurement of VBS signatures are performed 
separately using dedicated event selections. The resonance search focuses on high-invariant mass region of the 
diboson system, where the hadronically-decaying $V$ boson has large transverse momentum and is reconstructed 
as a single large-radius ($R$) jet. The VBS search considers a wide range of the diboson-system mass, exploiting the techniques 
to reconstruct the hadronic decay of the $V$ boson as either a large-radius jet or two small-radius jets.

\subsection{Resonance analysis}

Events are required to have exactly one electron or muon with $\pt>30$~GeV and $|\eta|<4.0$. 
Events are further required to contain a hadronically-decaying $W/Z$ candidate reconstructed as a 
large-radius jet~(denoted by $J$) with a distance parameter $R=1.0$, and a leptonically-decaying $W/Z$ candidate. 
The large-$R$ jet is required to have
 $\pt^J>200$~GeV and $|\eta_J|<2.0$.
If two or more large large-$R$ jets are found, the one with the highest \pt is chosen as a hadronically-decaying $W/Z$ boson candidate.
The missing transverse energy \met has to be greater than 80~GeV,  which suppresses multijet background to a negligible level.
By constraining the \met + lepton system to be consistent with the $W$-boson mass, the $z$ component of the neutrino momentum
can be reconstructed by solving a quadratic equation. The smallest solution is chosen and in the case where the solution is imaginary, only the real part is taken. 

The selected events are required to satisfy the mass-window cut of 
$|m_{J}-m_{W/Z}|<60$~GeV and  have $D_2< 1.5$, where $D_2$ is a jet substructure variable~\cite{fatjet_d2, BosonTagPaper} 
sensitive to a two-prong structure from hadronically-decaying $V$ boson.
If the selected large-$R$ jet contains a $b$-quark, then event is rejected 
to reduce contribution from $t\bar t$ background.

\begin{figure}
\centering

\includegraphics[width=0.45\textwidth]{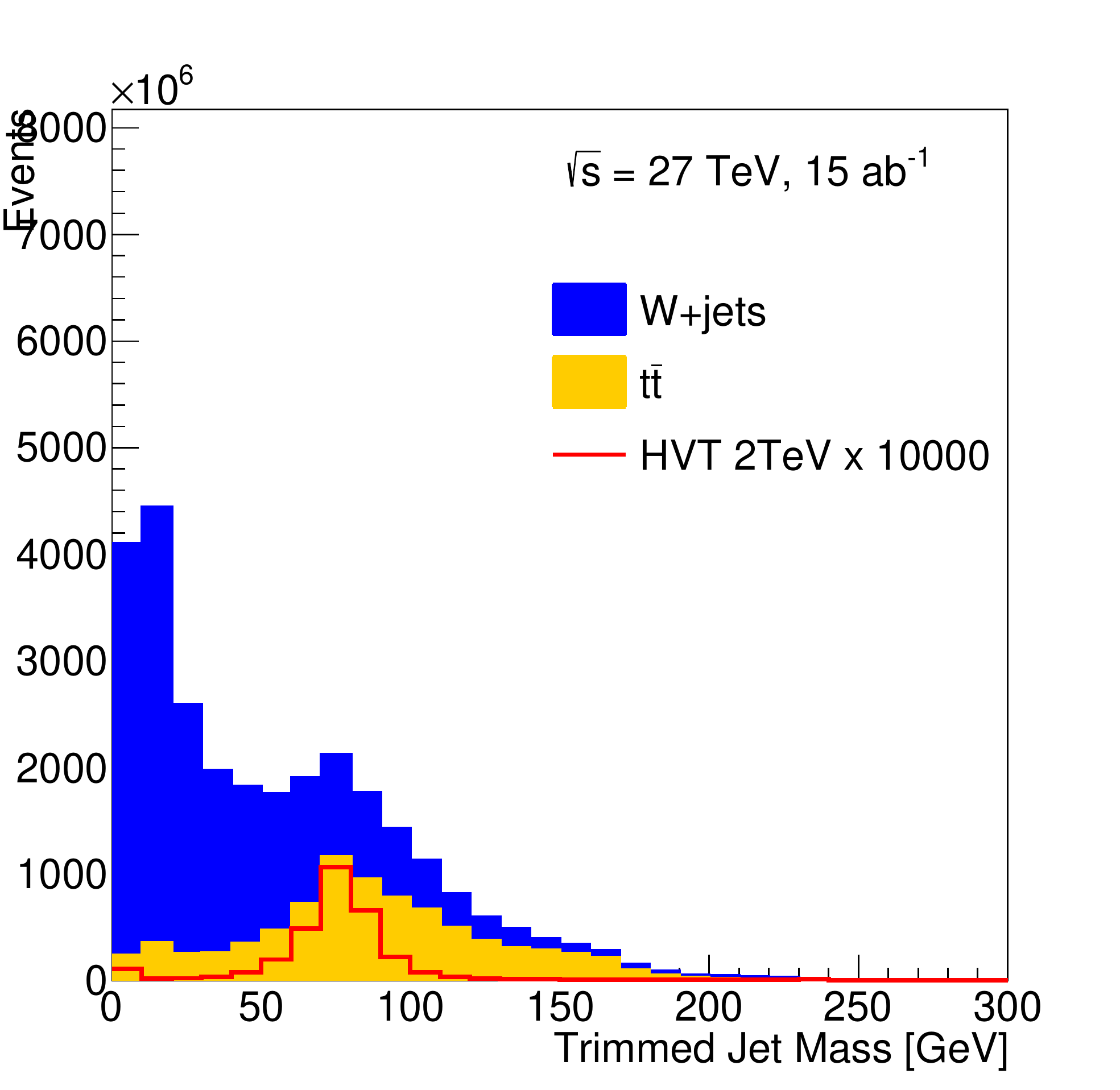}
\includegraphics[width=0.45\textwidth]{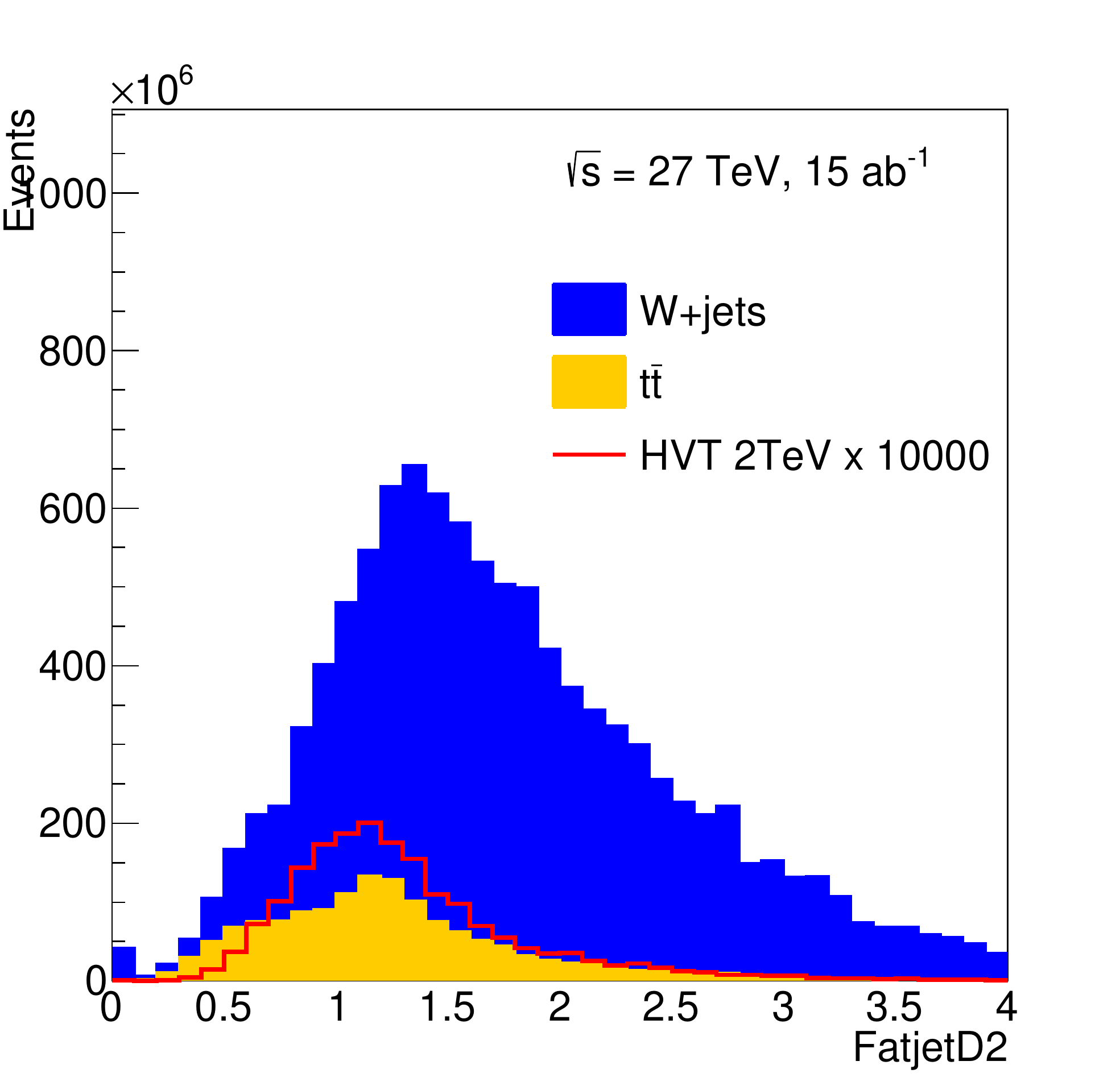}\\
\includegraphics[width=0.45\textwidth]{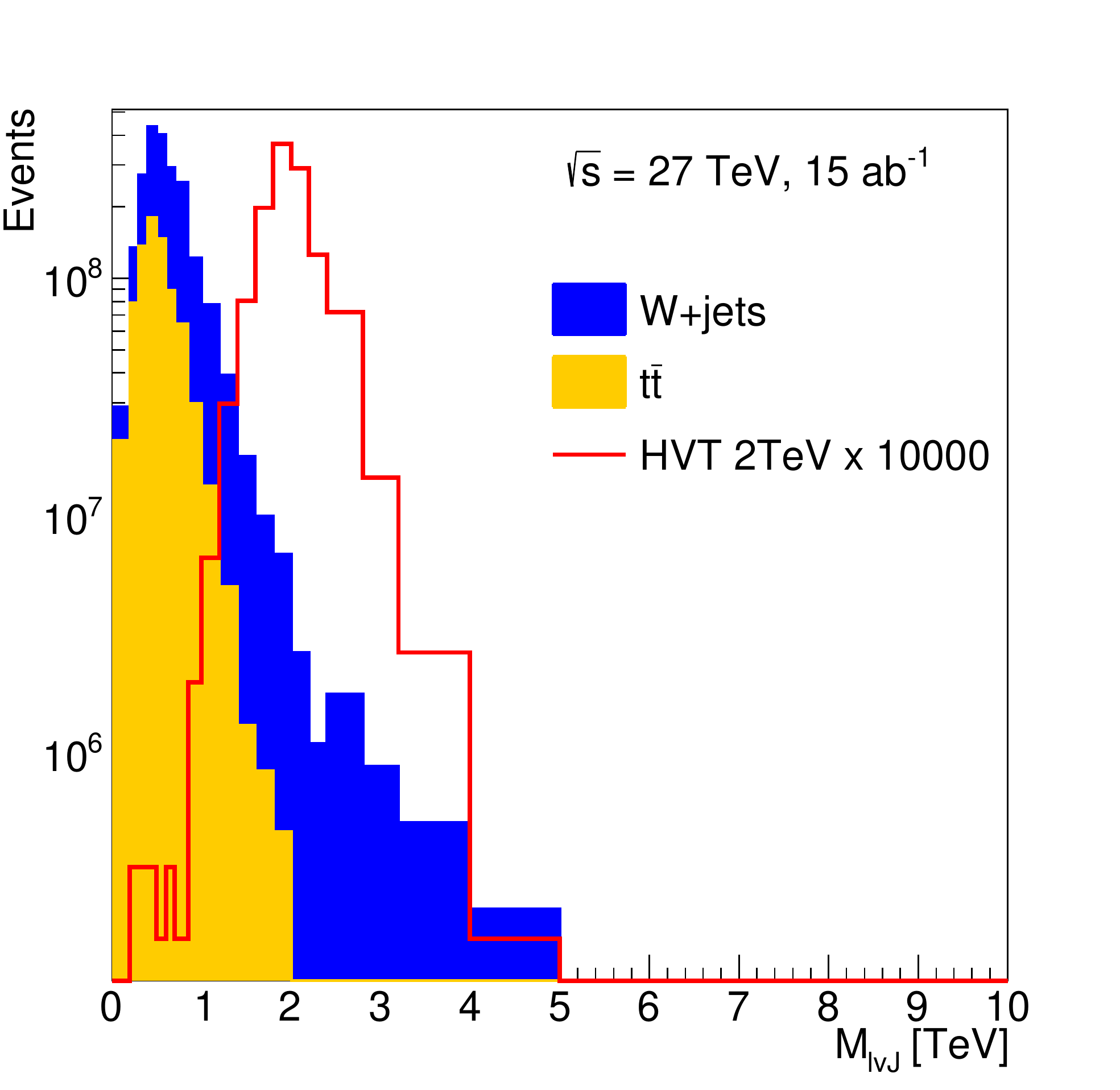}
\includegraphics[width=0.45\textwidth,height=0.29\textheight]{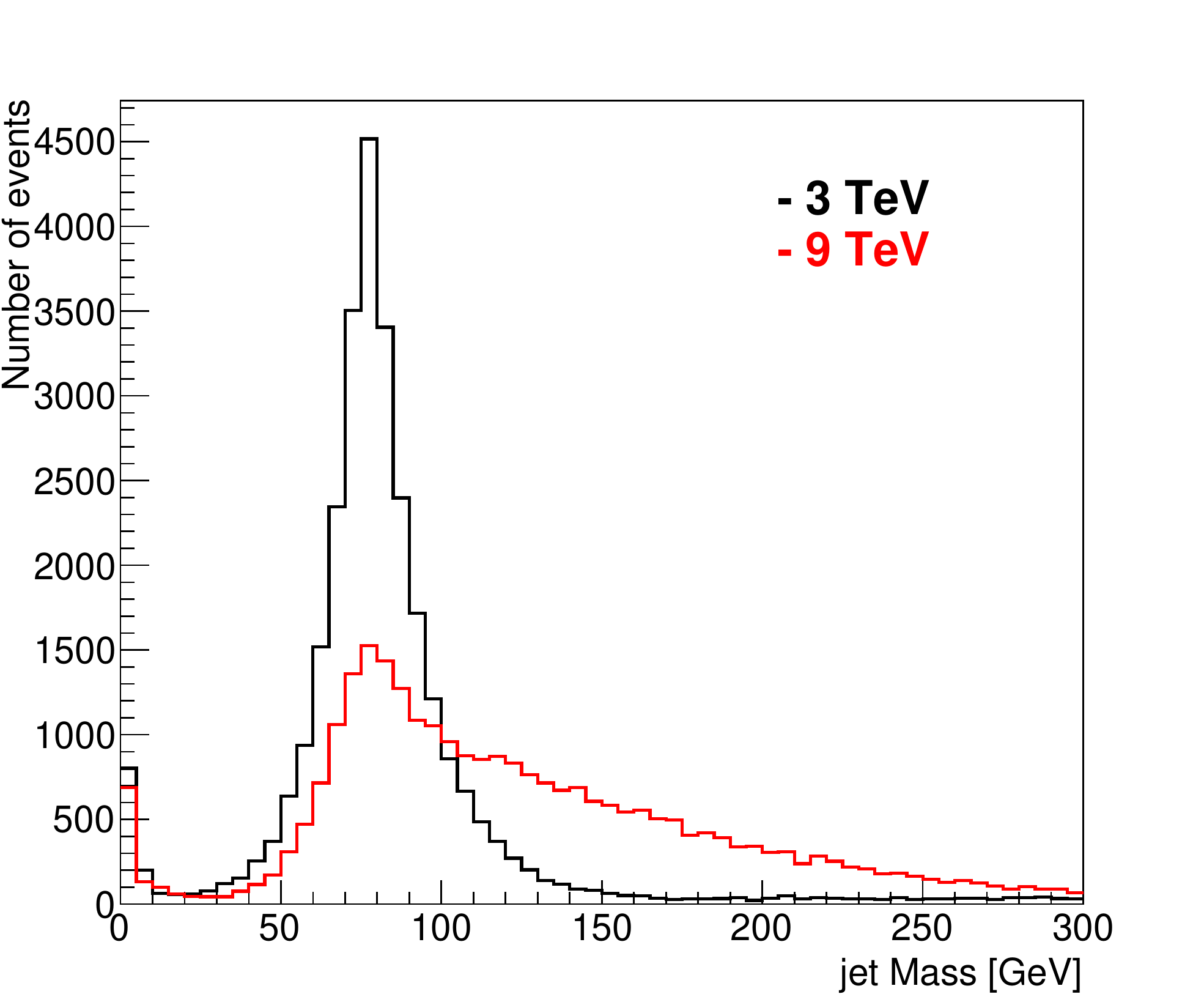}
\caption{Kinematic distributions of HVT signal and background ($W$+jets and $t\bar{t}$) events at 15~ab$^{-1}$ in the resonance analysis. The HVT signal with a 2~TeV resonance is shown after multiplying the expected yield by $10^4$.
The invariant mass of the diboson system $m_{lvJ}$ is used as a discriminant variable. The bottom-right plot shows jet mass distributions for the HVT signals with resonance masses of 3 and 9~TeV, 
illustrating the need of future improvement to mitigate the effect of energy merging within the 
limited detector granularity.}
\label{fig:KinematicReonance}
\end{figure}

The distributions of the relevant kinematic properties of the large-$R$ jets are shown in Fig.~\ref{fig:KinematicReonance}. The invariant mass of the diboson system ($m_{\ell\nu J}$) is most sensitive to resonant diboson production and hence used as a discriminant variable.

The background shape and normalization are constrained using dedicated control regions for systematic 
uncertainties associated with the background modeling. 
The following control regions are used in the final fit:
\begin{itemize}
\item If an event satisfies all the selection criteria except the $W/Z$-boson mass-window cut and has no $b$-jets ($b$-veto), then the event is categorized as a $W$  control region event.
\item If an event satisfies all the selection criteria and has additional $b$-jets outside the large-$R$ jet, then the event is categorized as a top  control region event.
\end{itemize}
These regions are used to constrain the $W$+jets  and top background normalization and shape uncertainties.

\subsection{VBS analysis}

Experimentally, VBS is characterized by the presence of a pair of vector bosons ($W$, $Z$, or $\gamma$) and two forward jets with large separation in pseudo-rapidity and a large invariant mass.
Therefore the VBS search requires events to have two additional forward jets characteristic of the VBS topology (called tagging jets) in addition to jets associated with the boson decay.

The tagging jets are required to be non-$b$-tagged in order to suppress contribution from diagrams with a $Wtb$ vertex
(especially the electroweak $t\bar{t}$ process) in the electroweak $VVjj$ production.
The tagging jets must be in the opposite hemispheres, $\eta(j_1^{\mathrm{tag}}) \cdot \eta(j_2^{\mathrm{tag}}) < 0$,
and have the highest dijet invariant mass among all pairs of jets remaining in the event after selecting jets from $V \to qq$ decay (called signal jets).  
After the tagging jet pair is selected, it is required that both tagging jets have $\pt>30$~GeV to suppress contribution from pile-up interactions, and that the invariant mass of the two tagging jets is greater than 200~GeV.
In the VBS search the tagging jets are selected after identifying the signal jets (as described below) and are required to be $\Delta R>1$ from the large-$R$ signal jets.

In contrast to the resonance analysis, the VBS analysis first selects events based on the large-$R$ jets using the same criteria as those in the resonance analysis (called merged selection), then apply the selection based on the small-$R$ jets (called resolved selection) if the event fails the merged selection. In the resolved selection events are required to have at least two small-$R$ jets with $R=0.4$, and the hadronically-decaying $W/Z$ candidate is reconstructed from the pair of small-$R$ jets with the mass $m_{jj}$ closest to the $W/Z$ mass among all combinations of jets with $\pt>20$~GeV. 
The leading (sub-leading) signal jet is further required to have $\pt>40$ (30)~GeV after the jet pairs are selected to improve separation between the signal and background. The invariant mass of the signal jets is then
required to fall in the mass window of $60<m_{jj}<110$~GeV.
To suppress backgrounds containing top quarks, events are vetoed if they contain any $b$-tagged jets away from the  hadronic $W/Z$ boson candidate by $\Delta R>1.0$.
In addition, if the $m_{jj}$ is within the $W$-mass window and both jets are $b$-tagged, the event is removed. This cut is not applied to the $Z$-boson candidate in order to retain signal with $Z\rightarrow b\bar{b}$ decay.

\begin{figure}
\centering
\includegraphics[width=0.45\textwidth]{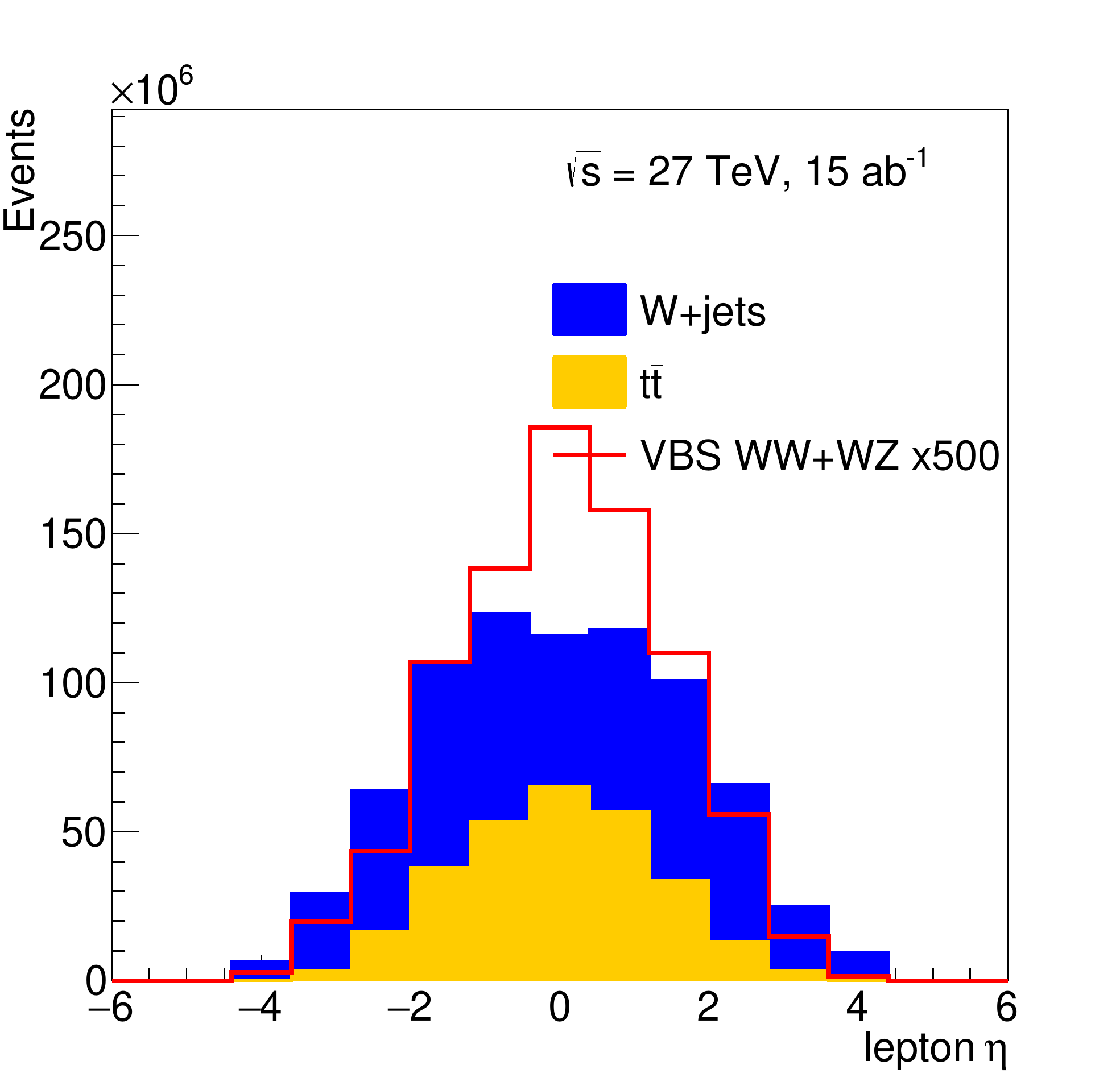}
\includegraphics[width=0.45\textwidth]{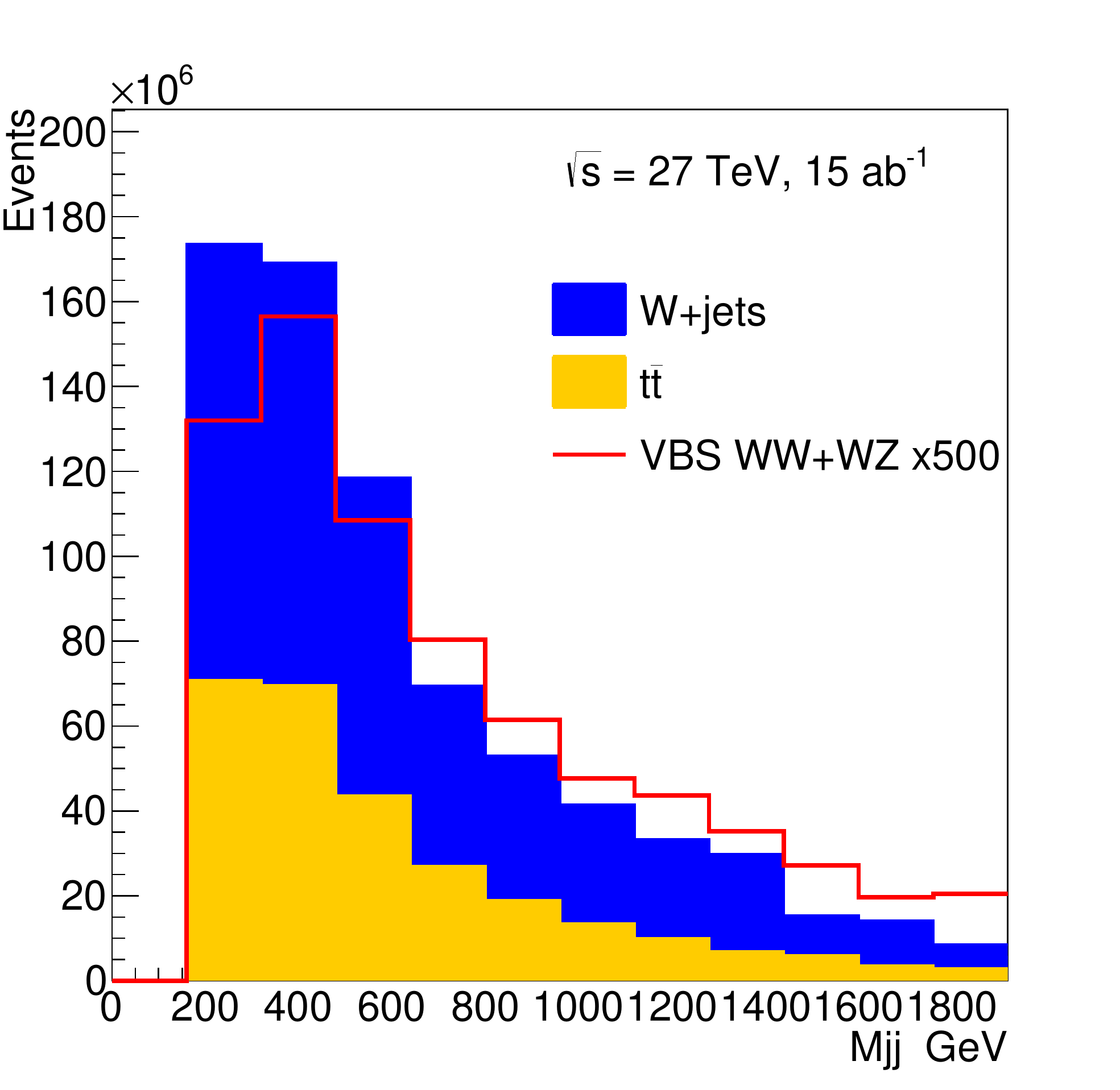}\\
\includegraphics[width=0.45\textwidth]{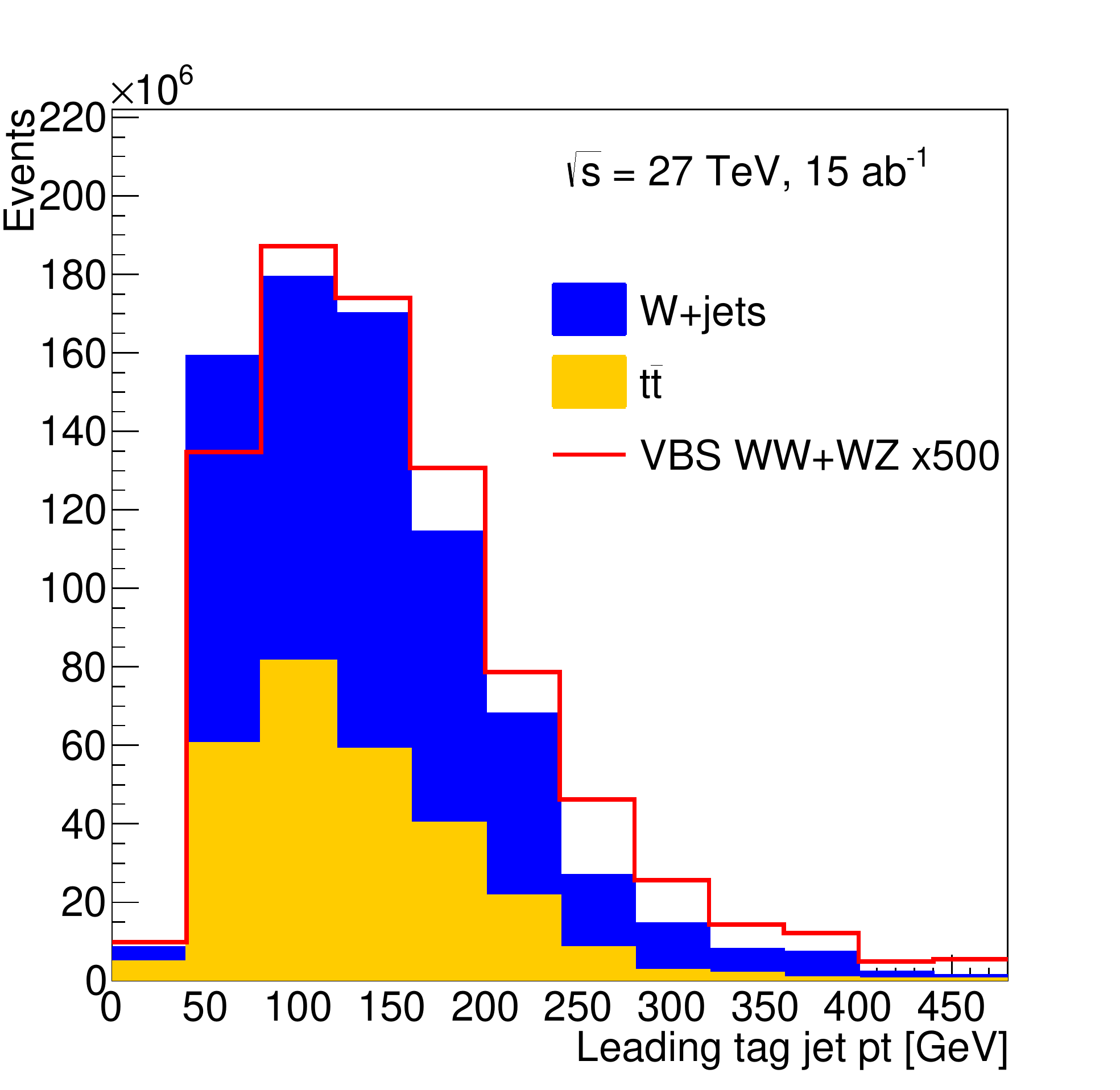}
\includegraphics[width=0.45\textwidth]{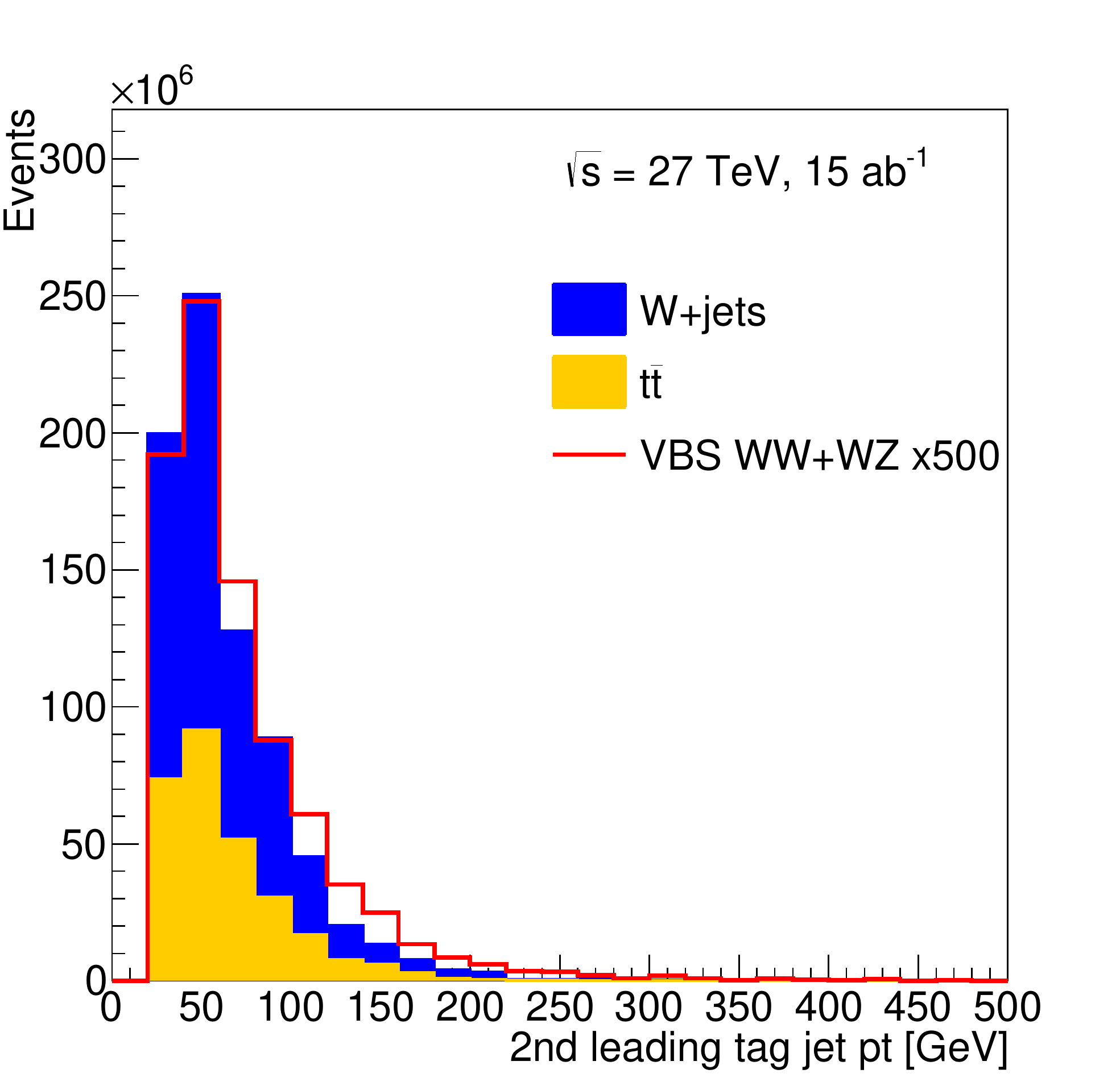}\\
\includegraphics[width=0.45\textwidth]{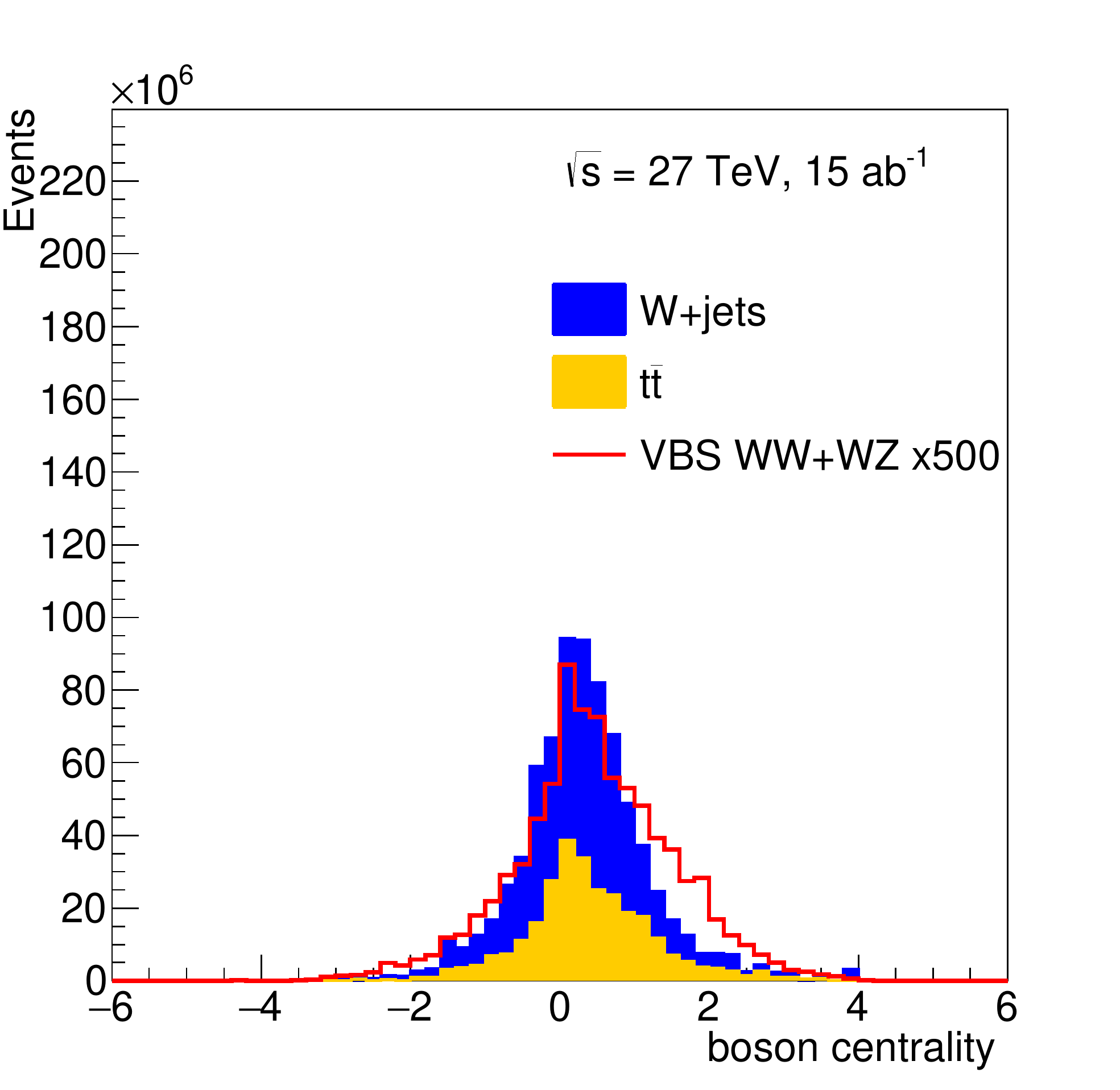}
\includegraphics[width=0.43\textwidth,height=0.26\textheight]{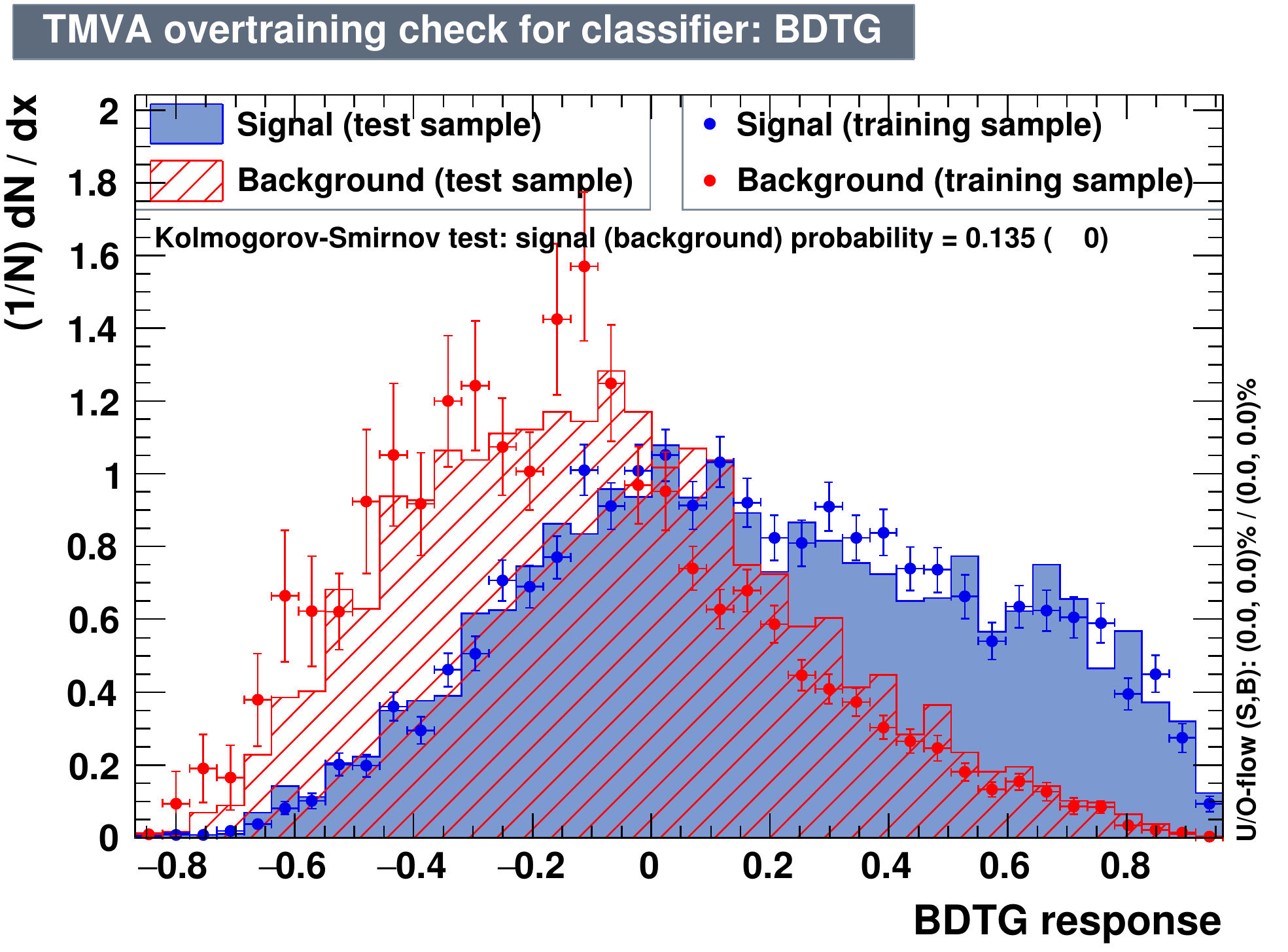}
\caption{Distributions of lepton $\eta$, invariant mass of the tagging jets, leading and sub-leading tagging jet \pt and boson centrality   for VBS signal and background ($W$+jets and $t\bar{t}$) events at 15~ab$^{-1}$ in the VBS analysis. The VBS signal with both $WW$ and $WZ$ processes is shown after multiplying the expected yields by 500. Shown for  the merged  selection. Bottom right distribution of the Boosted Decision Tree discriminant to separate VBS (electroweak $WWjj$) signal from $W$+jets and $t\bar{t}$ backgrounds in the VBS analysis.}
\label{fig:KinematicVBS1}
\end{figure}

To optimize the signal sensitivity, Boosted Decision Trees (BDT) are trained on the background and signal simulation samples separately for the resolved and merged selections.
Five variables are included in the merged and resolved BDT: the invariant mass of the diboson system 
($m_{\ell\nu J}$ for the merged and $m_{\ell\nu jj}$ for the resolved selections), the lepton $\eta$, 
the leading and sub-leading tagging jet \pt and the boson centrality $\zeta_V$. 
The boson centrality is defined as $\zeta_V=\text{min}(\Delta\eta_+,\Delta\eta_{-})$  where $\Delta\eta_{+}=\text{max}(\eta(j_1^{\mathrm{tag}}),\eta(j_2^{\mathrm{tag}}))-\text{max}(\eta(\ell\nu),\eta(J))$ and $\Delta\eta_{-}=\text{min}(\eta(\ell\nu),\eta(J))-\text{min}(\eta(j_1^{\mathrm{tag}}),\eta(j_2^{\mathrm{tag}}))$, as shown in Fig.~\ref{fig:KinematicVBS1}.
These variables are chosen as they are the minimal subset of variables with the greatest separation between the signal and background.
The BDT are trained using a gradient descent BDT algorithm, maximizing the Gini index, in the TMVA package~\cite{TMVA}.
The BDT output is chosen as a final discriminant and the distribution is used in the final fit for the VBS search.
Similarly to the resonance search, if an event fails either the $W/Z$ mass-window cut or the $b$-veto requirement but passes all the other selection criteria, then the event is categorized into a $W$- or top-control region. 
The BDT response used for the VBS search is shown  in Fig.~\ref{fig:KinematicVBS1}.

\section{Pileup effects on boson identification}
%
The unprecedented energy of $pp$ collisions at the HE-LHC will significantly improve sensitivity to new multi-TeV particles over LHC and HL-LHC. However, the experimental environment is expected to be challenging at the HE-LHC, primarily due to a significant increase of the number of $pp$ collisions in a same and nearby bunch crossings (pile-up). The HE-LHC is planned to be operated at a centre-of-mass energy of 27~TeV with a beam intensity of about $2\times10^{11}$ protons per bunch, resulting in a peak luminosity of about $3\times10^{35}$~cm$^{-2}$s$^{-1}$ and 800 pile-up collisions at the peak luminosity. Such extreme pile-up condition is expected to be particularly challenging for identifying hadronically-decaying $W$/$Z$ boson as the extra contribution of particles produced from pile-up collisions into jets could degrade the performance of $W$/$Z$-boson tagger significantly. It is therefore important to assess the performance of pile-up mitigation technique at the HE-LHC in order to have a reliable estimate of the search sensitivity.

The study presented here focuses on the performance of pile-up mitigation techniques and $W$/$Z$-boson tagging. The VBS-signal events are produced with the overlay of minimum-bias $pp$ interactions generated using Pythia~8.219. The minimum-bias interactions are overlaid onto hard scattering event using Poisson probability distribution with the mean number of interactions ($\mu_{\text{pileup}}$) varied from 0 to 100, 200, 400 and 800. Furthermore, the minimum-bias interactions are distributed randomly in $z$ and timing using Gaussian profiles of $\sigma_z=5.3$~cm and $\sigma_\text{t}=160$~ps, respectively ($z$=0 at the detector center and t=0 for hard scattering event). The overlaid VBS-signal events are processed through Delphes with two pile-up mitigation techniques: the Pile-up Per Particle Identification (PUPPI) algorithm~\cite{Bertolini:2014bba} used in CMS and the trimming procedure used in ATLAS. The trimming parameters of the \pt fraction cut and the sub-jet reclustering radius are chosen to be the same as those used in ATLAS. For the PUPPI algorithm the standard Delphes implementation is used. The Delphes detector configuration is identical regardless of whether the pile-up overlay is included or not. 

\begin{figure}
\centering
\includegraphics[width=0.45\textwidth]{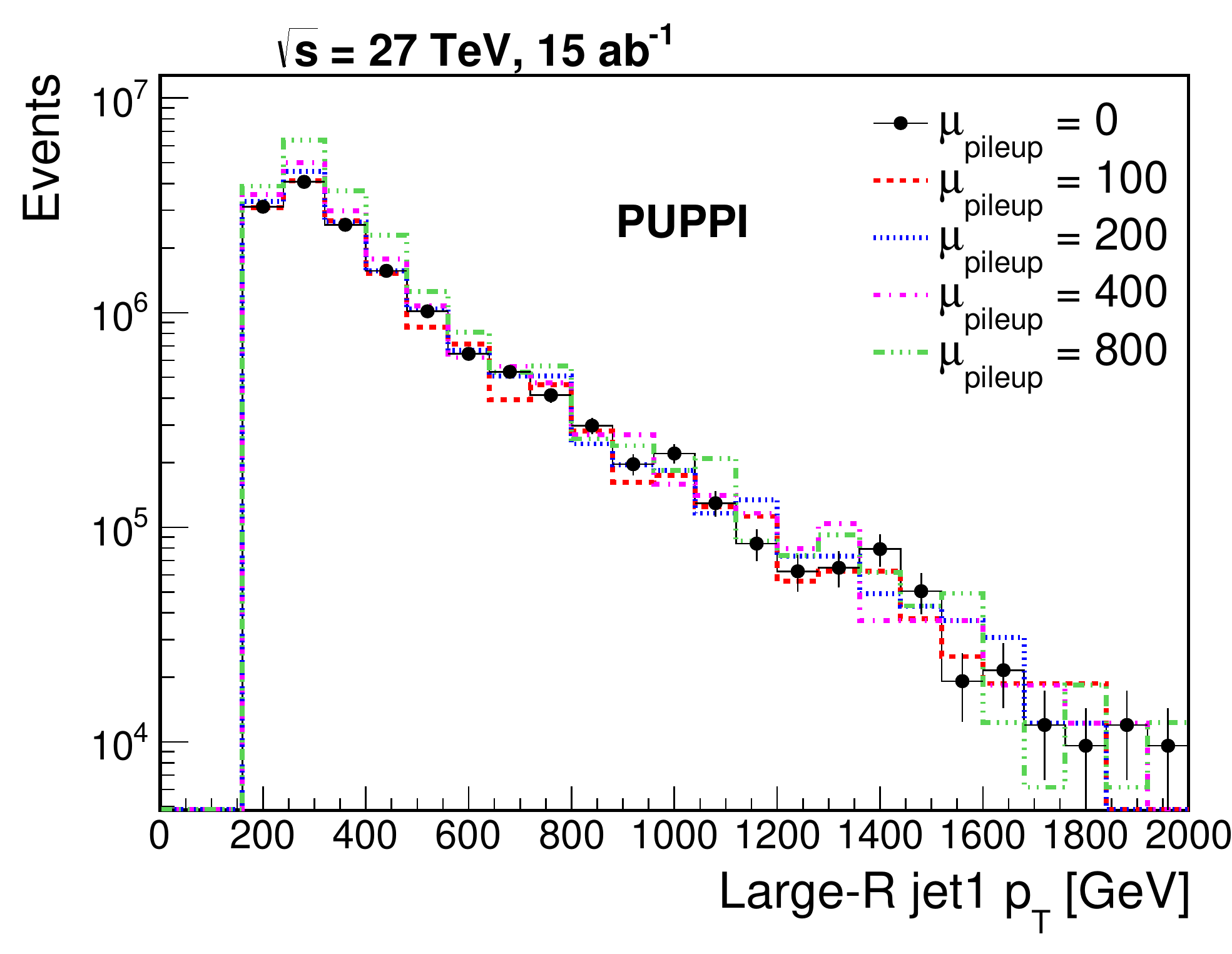}
\includegraphics[width=0.45\textwidth]{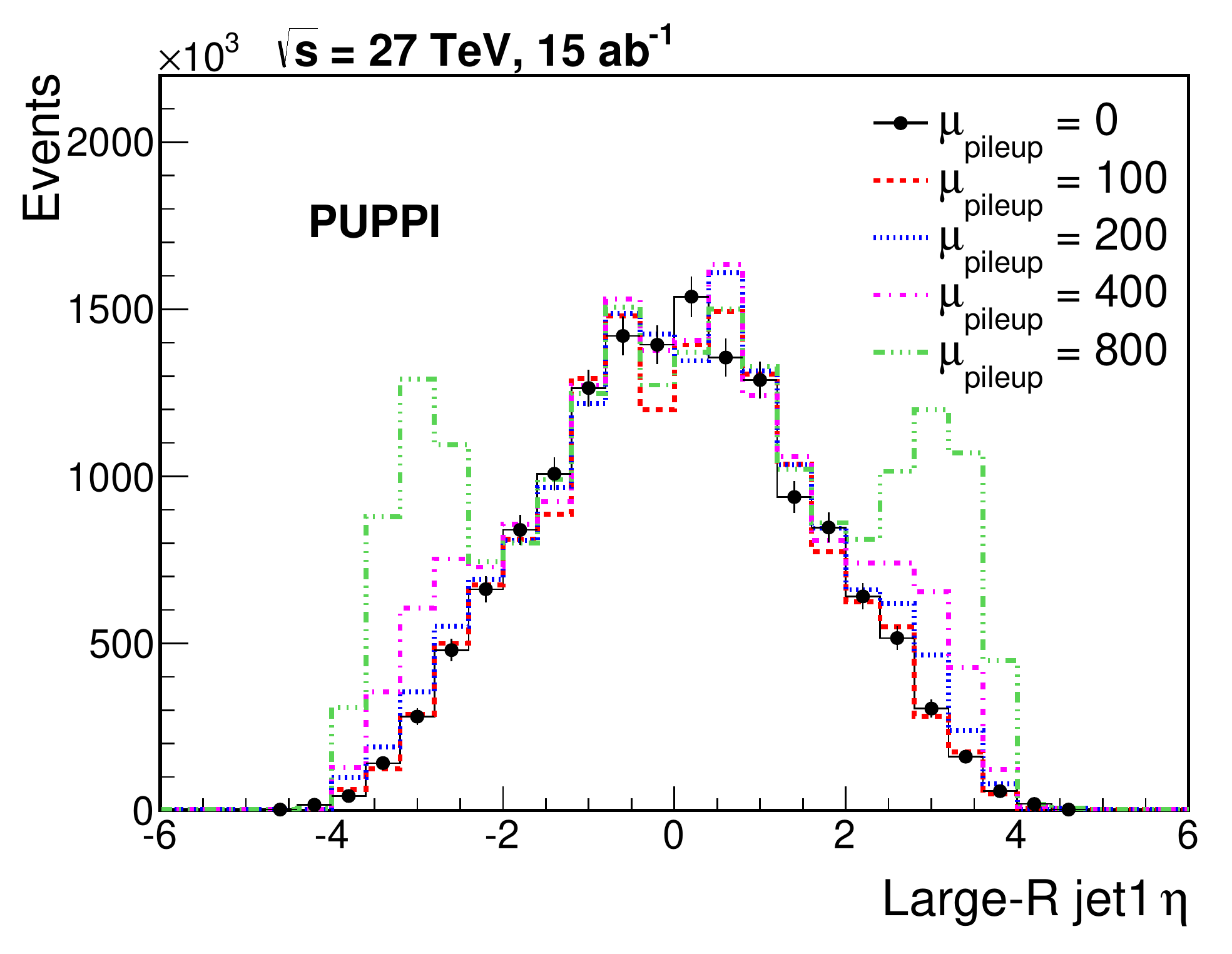}
\caption{Leading large-$R$ jet \pt (left) and $\eta$ (right) distributions after applying the PUPPI algorithm 
at an integrated luminosity of 15~ab$^{-1}$ at $\sqrt{s}=27$~TeV with five different pile-up overlay conditions of 
$\mu_{\text{pileup}}=0$, 100, 200, 400 and 800.}
\label{fig:PUPPI_jet}
\end{figure}

\begin{figure}
\centering
\includegraphics[width=0.45\textwidth]{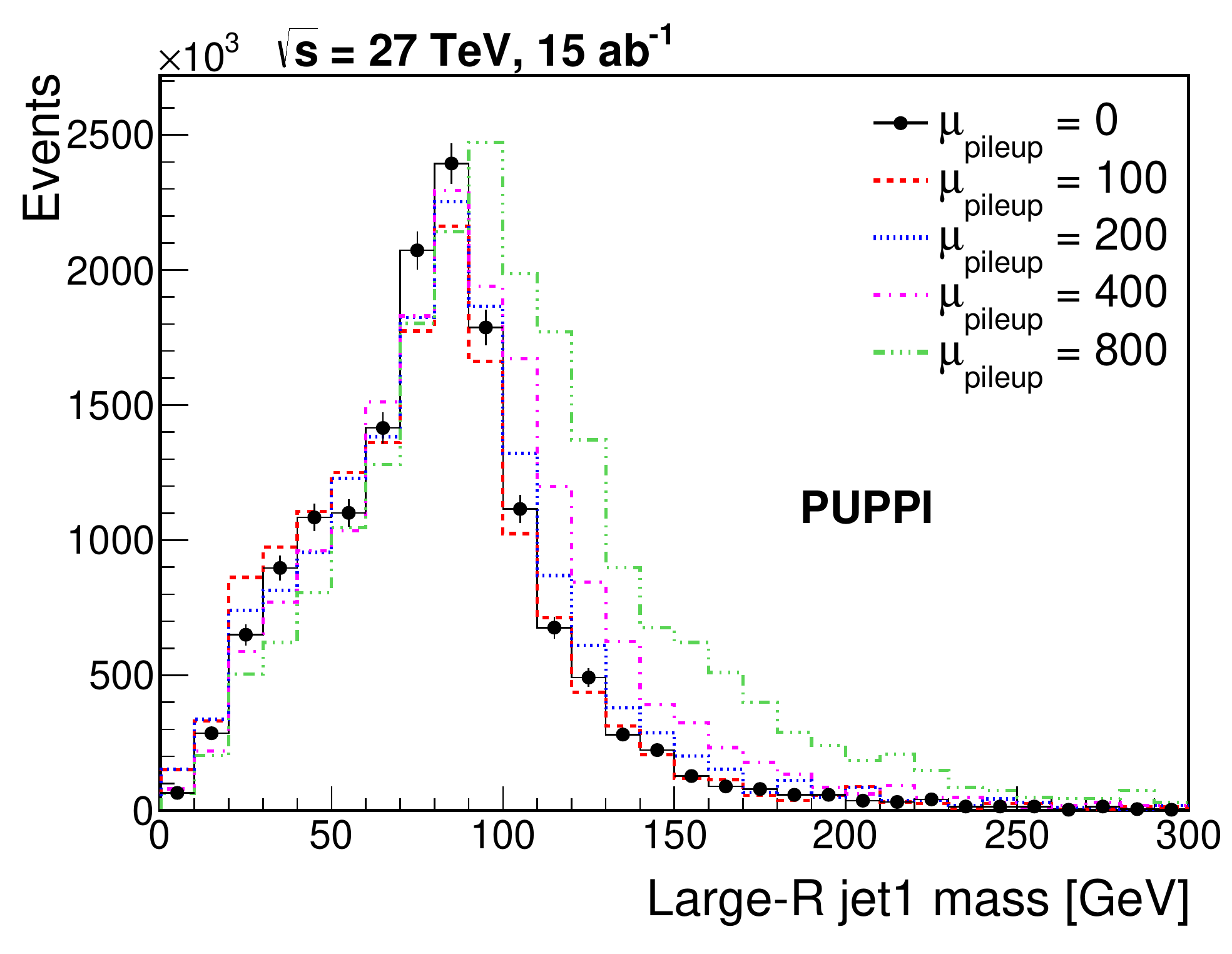}
\includegraphics[width=0.45\textwidth]{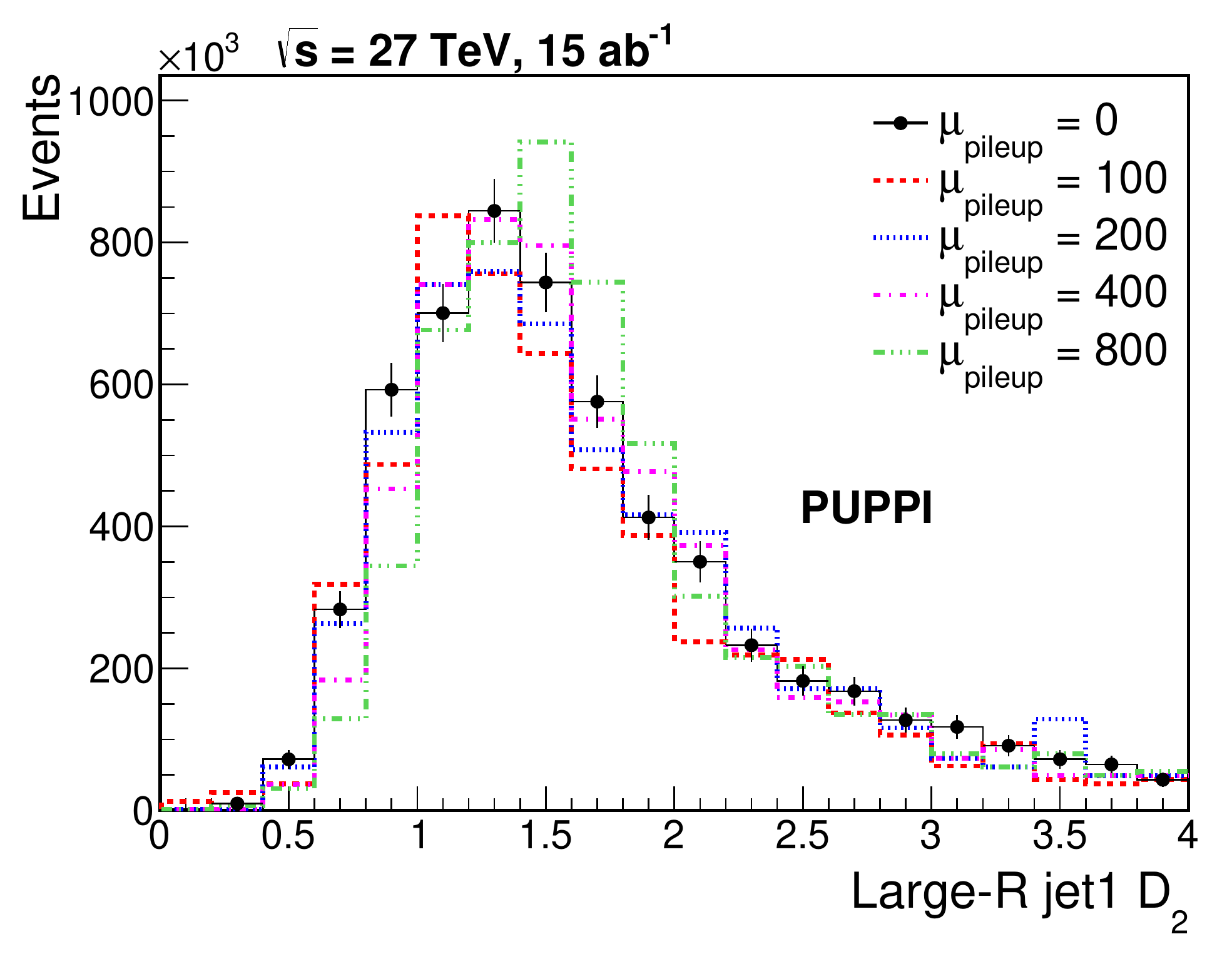}\\
\includegraphics[width=0.45\textwidth]{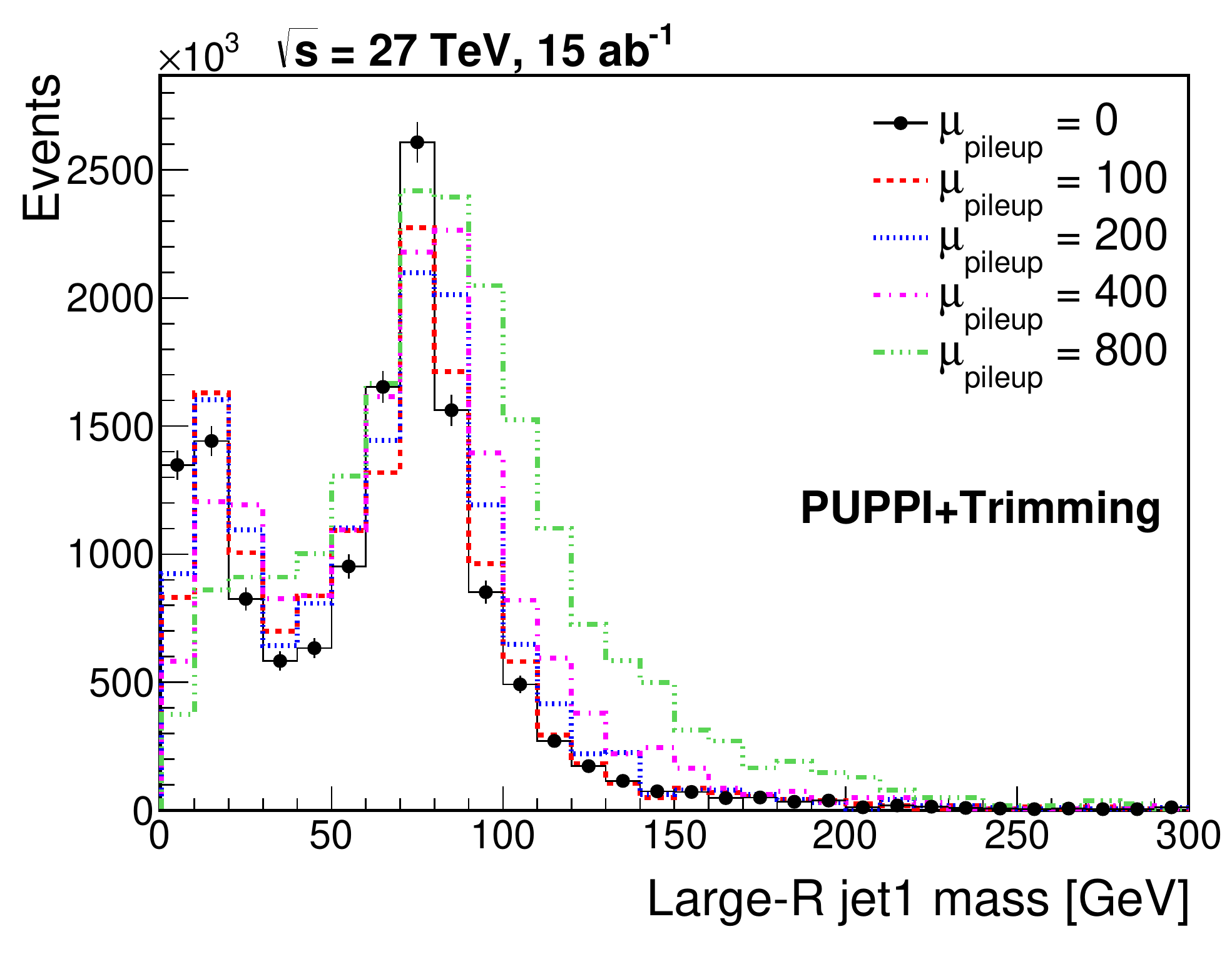}
\includegraphics[width=0.45\textwidth]{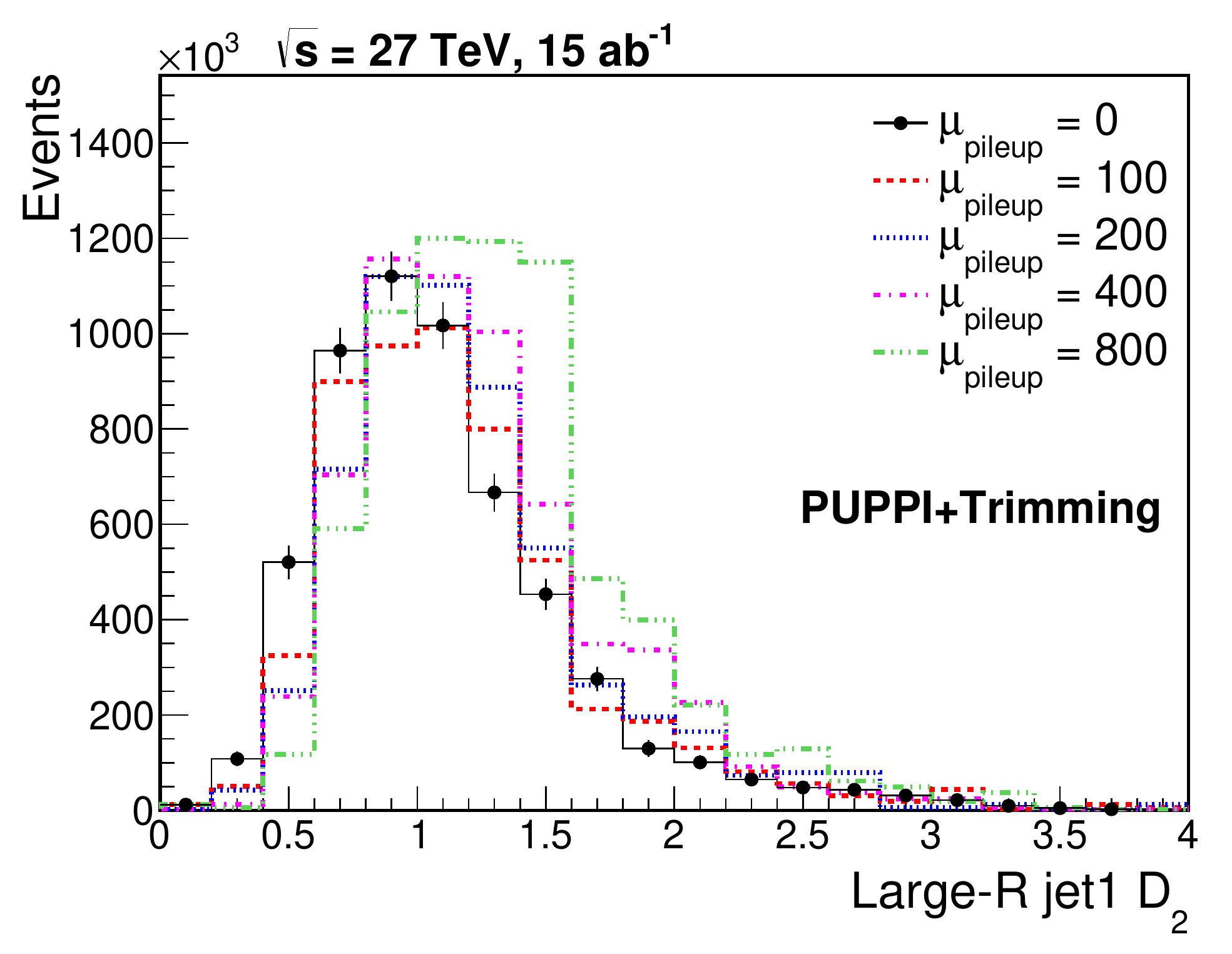}
\caption{Leading large-$R$ jet mass (top-left) and $D_2$ (top-right) distributions with $\pt>200$~GeV 
after applying the PUPPI algorithm at an integrated luminosity of 15~ab$^{-1}$ at $\sqrt{s}=27$~TeV with 
five different pile-up overlay conditions of $\mu_{\text{pileup}}=0$, 100, 200, 400 and 800. The bottom two plots show the same
distributions but after additionally requiring that the jets are trimmed with the condition described in the text.}
\label{fig:PUPPI_trimmed_jet}
\end{figure}

Figure~\ref{fig:PUPPI_jet} shows the \pt and $\eta$ of the leading large-$R$ jet with $\pt>200$~GeV in VBS events with the five pile-up overlay conditions. The jets are processed with the PUPPI algorithm but not the trimming in the figure. The jet \pt spectra are similar while the $\eta$ distributions show increasing jet rate at large $|\eta|$ above 2 when the $\mu_{\text{pileup}}$ is 200 or higher. This behaviour is not significantly altered even when the trimming procedure is applied in addition to PUPPI. Figure~\ref{fig:PUPPI_trimmed_jet} shows the leading large-$R$ jet mass ($m_J$) and $D_2$ distributions for the PUPPI-only jets and the PUPPI+trimmed jets, both required to have $\pt>200$~GeV. The mass-window cut of $|m_J-m_{W/Z}|<15$~GeV is applied for events used in the $D_2$ distributions. Both $m_J$ and $D_2$ distributions get shifted towards lower values with the trimming applied, enhancing the peak around $m_W$ and the characteristics of two-prong structure for the $W$-boson decay. The residual pile-up effect is still visible as a shift towards larger values with increasing $\mu_{\text{pileup}}$, but the overall signal yield after the mass-window and $D_2$ requirements (e.g, $D_2<1.5$) is largely stable. This indicates that an impact to the $W$/$Z$-boson tagging performance from expected pile-up collisions at the HE-LHC can be mitigated to the level where the tagging performance at the HE-LHC is similar to that at Run 2 or HL-LHC. Therefore, the study presented in the rest of this paper is based on the $W$/$Z$-boson tagging performance at Run 2.

\section{Systematic uncertainties}
%
Two categories of systematic uncertainties are considered in the analysis: experimental uncertainties associated with the detector response and calibration of reconstructed objects, and uncertainties on the background modeling. Among those uncertainties, most dominant systematic sources are considered in each of the two categories. 

For experimental sources, the jet energy scale and resolution uncertainties are considered for the small (large)-$R$ jets used in the resolved VBS selection (merged resonance and VBS selections). In addition, the jet mass scale and resolution uncertainties are considered for the merged selections. The scale and resolution uncertainties are considered to be 10\%. 

For theoretical uncertainties, the normalization uncertainties are considered for the major backgrounds, i.e, $\pm10$\% for $W$+jets and $t\bar{t}$ backgrounds. The uncertainties on the background distribution shapes are taken into account for $W$+jets and $t\bar{t}$ processes by taking the variation of the diboson mass and BDT score distributions from different MC generators in Run 2 analysis. The uncertainties due to limited statistical accuracy of simulation predictions are not considered in the analysis.
 
\section{Extraction of analysis sensitivities}
%
The analysis sensitivities to the resonance and VBS signals are extracted by performing a simultaneous binned maximum-likelihood fit to the $m_{\ell\nu J}$ or $m_{\ell\nu jj}$ (BDT) distributions for the resonance (VBS) analysis in the signal regions and the $W$+jets and $t \bar{t}$ control regions. 
A test statistic based on the profile likelihood ratio~\cite{Cowan:2010js} is used to test hypothesized values of the global signal-strength factor ($\mu$), separately for each signal model considered. The likelihood is defined as the product of the Poisson likelihoods for all signal and control regions for a given signal category. The fit includes two main background contributions from $W$+jets and $t \bar{t}$. The $W$+jets and $t \bar{t}$ backgrounds are constrained by the corresponding control regions and are treated as uncorrelated among the resolved and merged signal regions.

Systematic uncertainties are taken into account as constrained nuisance parameters with Gaussian or log-normal distributions. For each source of systematic uncertainty, the correlations across bins in the $m_{\ell\nu J}$ and $m_{\ell\nu jj}$ distributions and between different kinematic regions as well as those between signal and background are taken into account.

\subsection{Resonance analysis}
\begin{figure}
\centering
\includegraphics[width=0.45\textwidth]{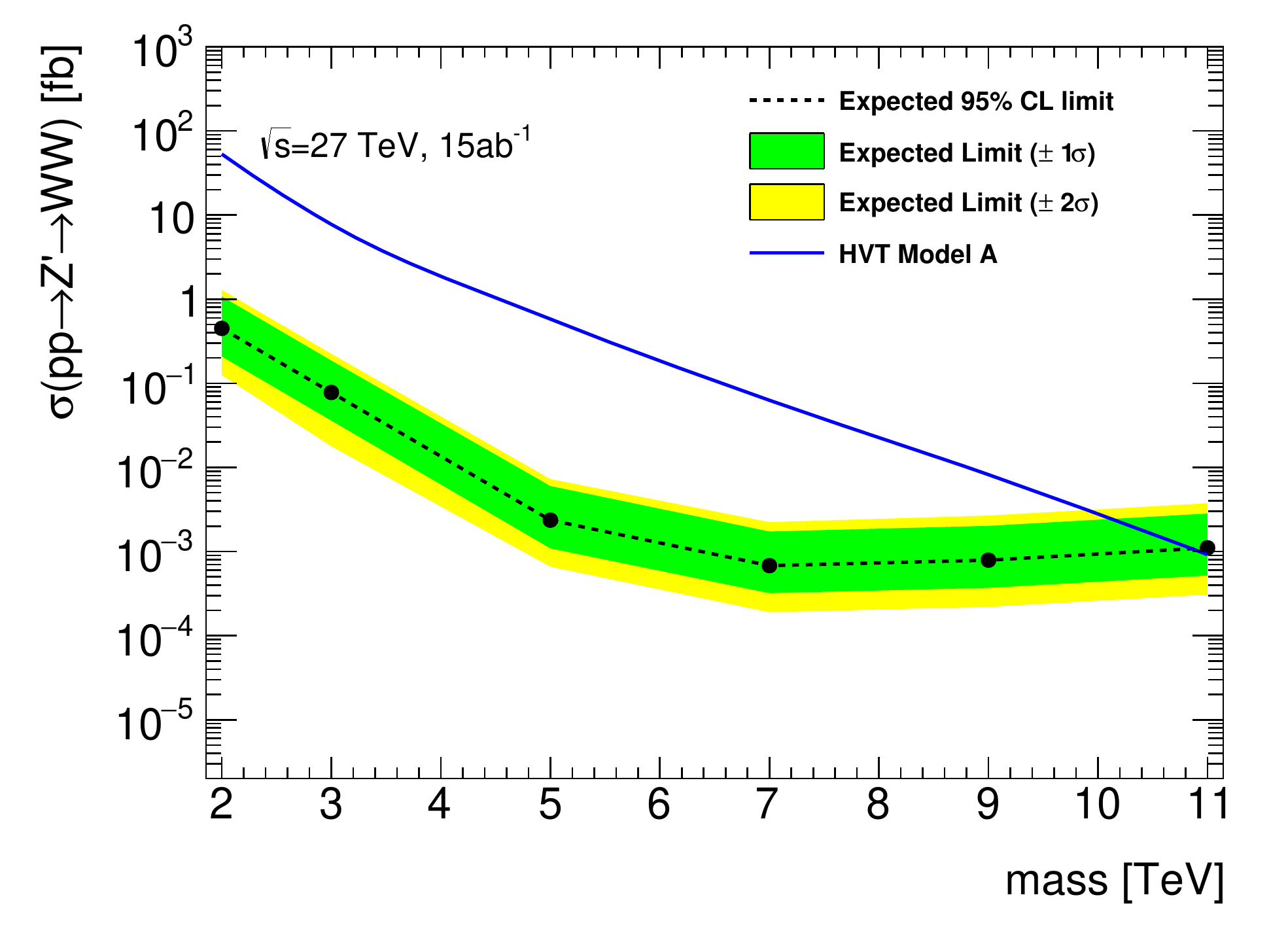}
\includegraphics[width=0.45\textwidth]{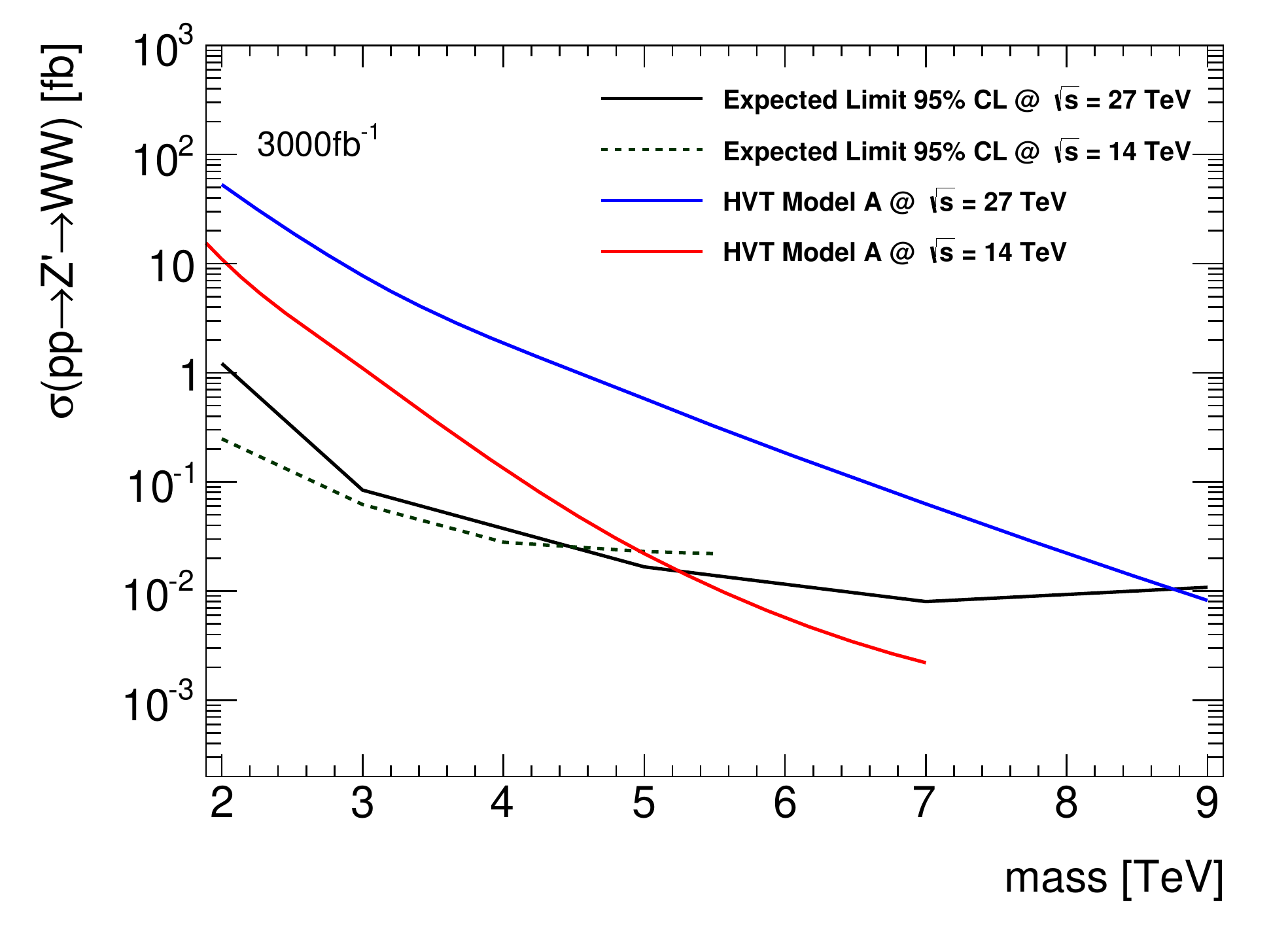}
\caption{95\% CL upper limits on the $Z'$ signal production cross section times branching fraction to $WW$ for 
the HVT model A with integrated luminosities of 15 (left) and 3~ab$^{-1}$ (right) at $\sqrt{s}=27$~TeV. The expected limits at the HL-LHC with 3~ab$^{-1}$ at $\sqrt{s}=14$~TeV are also shown in the right figure.}
\label{fig:Limits}
\end{figure}

A statistical analysis is performed using the CLs method to determine the expected upper limits 
on the signal cross section in the absence of signal. 
The $m_{\ell\nu J}$ distribution is constructed in the resonance analysis for
the background-only hypothesis and for a signal-plus-background hypothesis with varying signal strength
$\mu$, where $\mu=0$ corresponds to the background-only model and $\mu=1$ is the prediction of the signal model.
A likelihood function based on the binned $m_{\ell\nu J}$ distribution is used to exclude values of $\mu$ with 95\%
confidence. The expected upper limit set on the signal cross section is the greatest value of $\mu$ that is
not excluded with 95\% confidence. This procedure is repeated for each hypothetical signal mass. 
The expected upper limits set on the HVT $Z'$ signal cross section times branching fraction to $WW$ as a function of the $Z'$ mass are shown in Figure~\ref{fig:Limits} for the integrated luminosities of 3 and 15~ab$^{-1}$ at $\sqrt{s}=27$~TeV. 
Based on the $Z'$ production cross section from the HVT signal model the exclusion mass reach is extracted to be 9 and 11~TeV for
3 and 15~ab$^{-1}$. Also shown in Fig.~\ref{fig:Limits} is the sensitivity expected at the HL-LHC with 3~ab$^{-1}$ at $\sqrt{s}=14$~TeV~\cite{HLLHC_PUBnote}. A nearly factor 2 improvement in the mass reach is foreseen at the HE-LHC.

\begin{figure}
\centering
\includegraphics[width=0.5\textwidth]{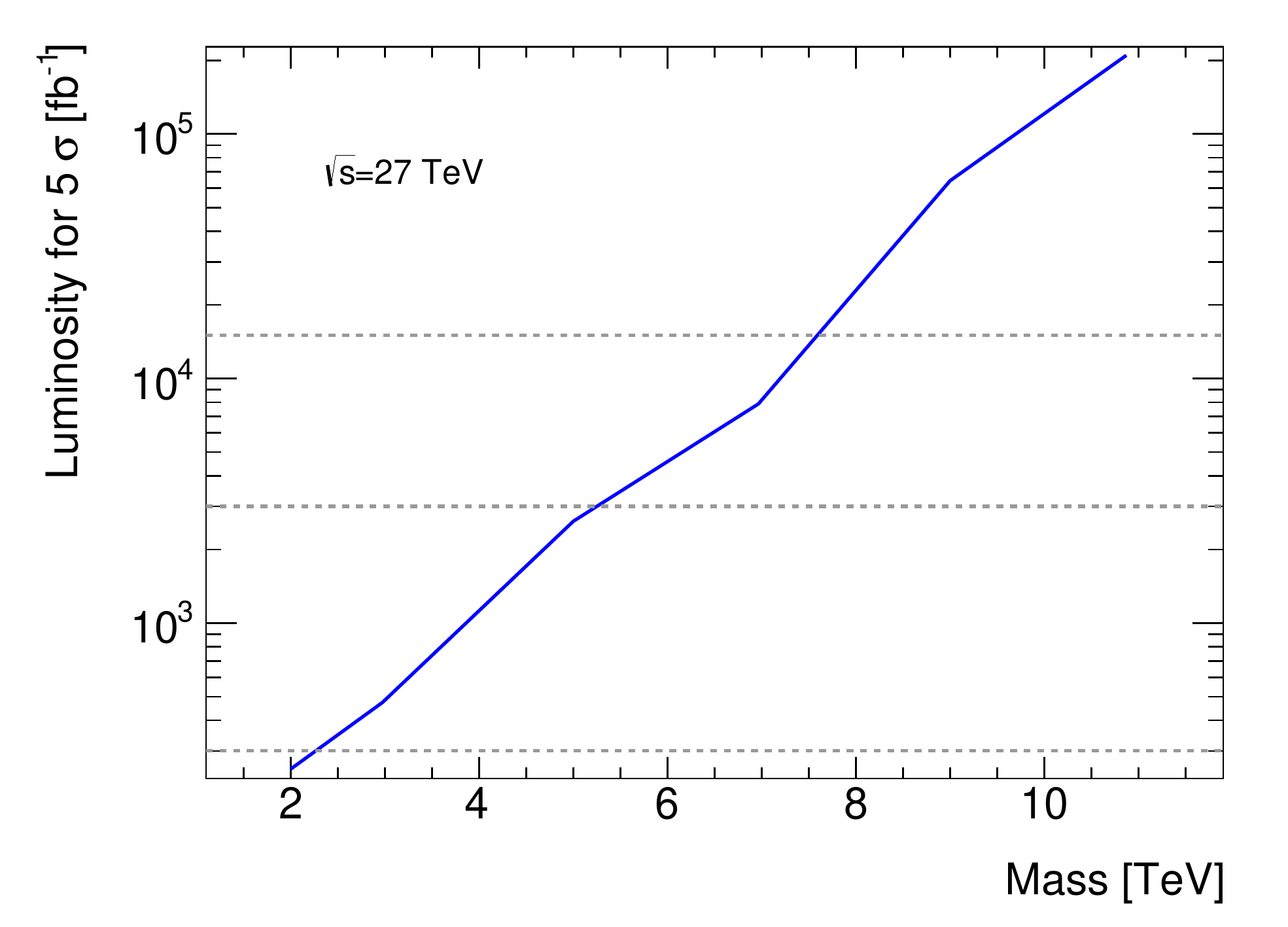}
\caption{Expected luminosity required to observe HVT $Z'$ signal with a $5\sigma$ significance as a function of the $Z'$ mass at $\sqrt{s}=27$~TeV.}
\label{fig:ResonanceDiscoverySignificance}
\end{figure}

The discovery sensitivity is also extracted as the luminosity required to achieve a $5\sigma$ observation of the signal. 
Here the signal significance is defined as the square sum of $s/\sqrt{s+b}$ calculated over the binned $m_{\ell\nu J}$ 
distributions at a given luminosity, where $s$ ($b$) is the signal (background) yield in each $m_{\ell\nu J}$ bin. 
Figure~\ref{fig:ResonanceDiscoverySignificance} shows the expected discovery significance 
as a function of integrated luminosity for the resonance search. With 15~fb$^{-1}$ of data the mass reach is extended to about 8~TeV, more than a factor 2 increase from the sensitivity expected at the HL-LHC with 3~ab$^{-1}$ at $\sqrt{s}=14$~TeV~\cite{HLLHC_PUBnote}.

\subsection{VBS Analysis}
For the VBS search, the statistical analysis is performed using the BDT distributions and the upper limits on the signal strength for the SM VBS~($WW/WZ\to \ell\nu qq$) processes are extracted. 

\begin{figure}
\centering
\includegraphics[width=0.45\textwidth]{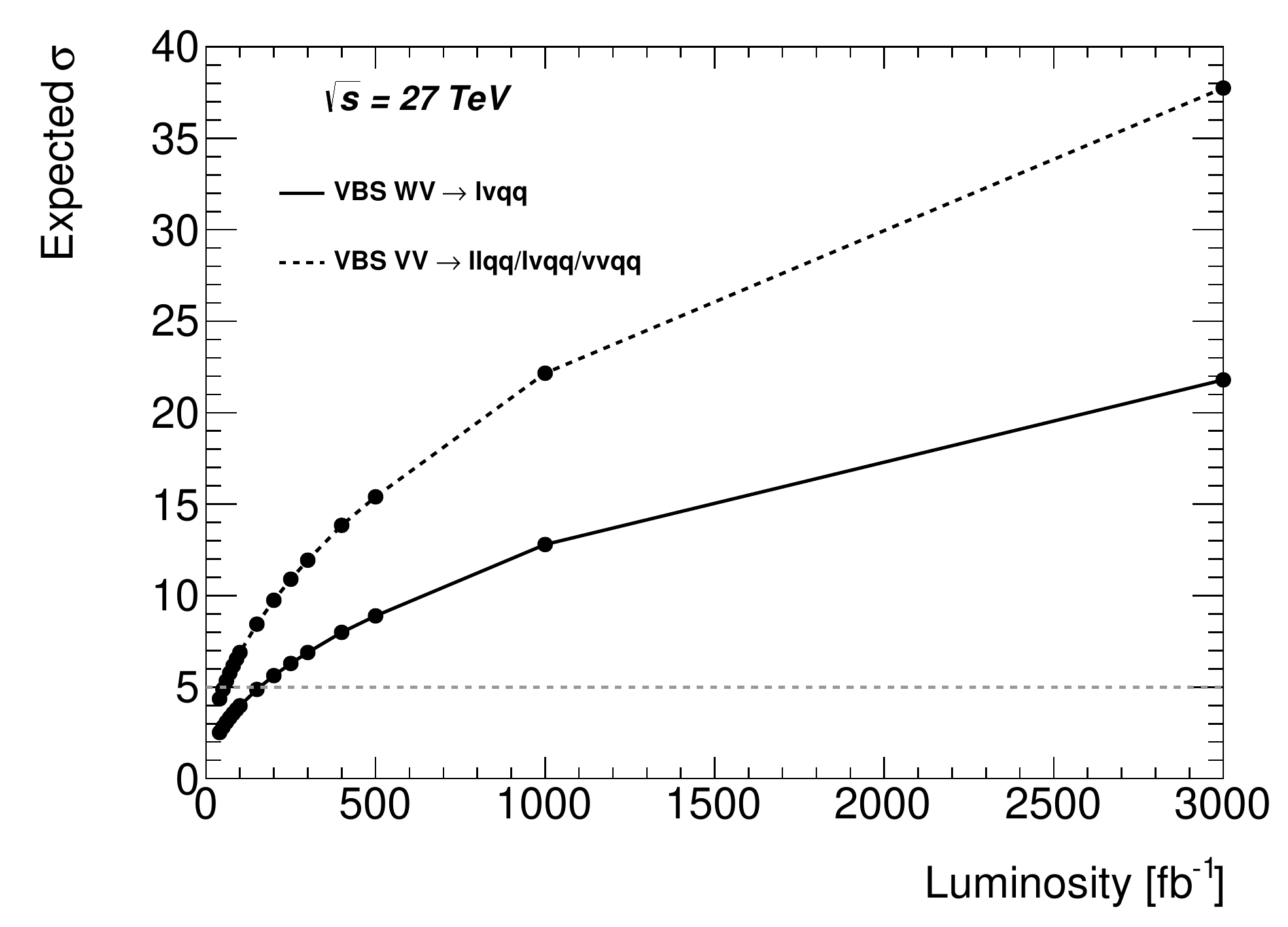}
\includegraphics[width=0.45\textwidth]{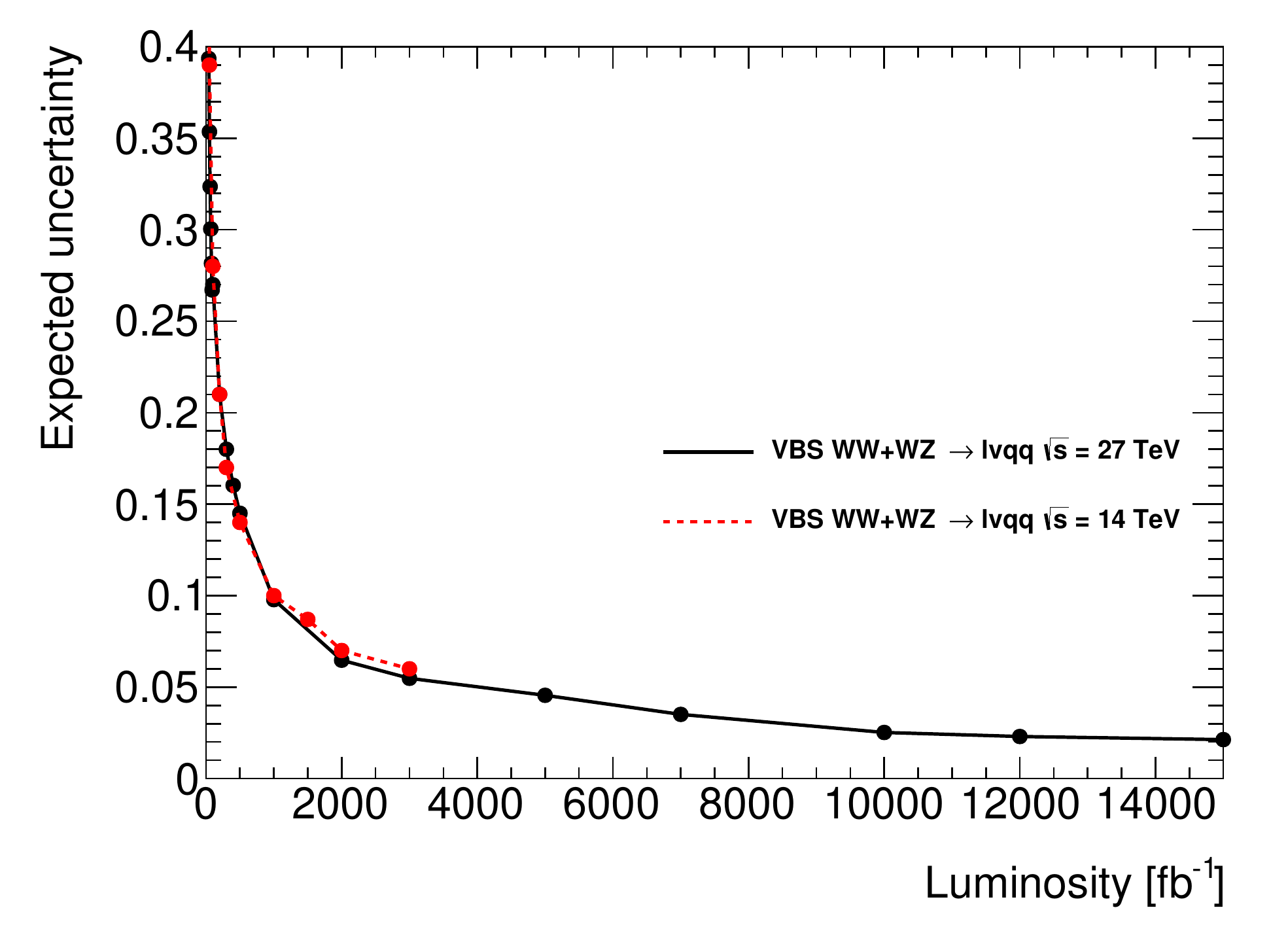}
\caption{Observed significance as a function of integrated luminosity (left) and expected cross-section uncertainty (right) for the VBS signal in the $\ell\nu qq$ channel at $\sqrt{s}=27$~TeV. The dashed line on the left shows expected significance from the combination of all the three semi-leptonic channels assumed to have sensitivity similar to the $\ell\nu qq$ channel. Also shown on the right is the expected cross-section uncertainty from the HL-LHC at $\sqrt{s}=14$~TeV.
}
\label{fig:VBSDiscoveryReach}
\end{figure}

The expected significance for the SM VBS processes is $\sim$ 7$\sigma$ at 300~fb$^{-1}$. The expected cross-section uncertainties are 18\% at  300~fb$^{-1}$ and 6.5\% at 3~ab$^{-1}$. The effects from the detector reconstruction efficiency and migration due to detector resolution are not considered for the estimates of cross-section uncertainties.
Figure~\ref{fig:VBSDiscoveryReach} shows the expected signal sensitivity and cross-section uncertainty as a function of integrated luminosity. 
In addition to the $\ell\nu qq$ channel, curves representing the estimated combined sensitivity including the other semi-leptonic channels, $\ell\ell qq$ and $\nu\nu qq$, are shown assuming that they have equal sensitivity as the $\ell\nu qq$ channel.

\subsection{Longitudinally polarized signal}
\begin{figure}
\centering
\includegraphics[width=0.4\textwidth]{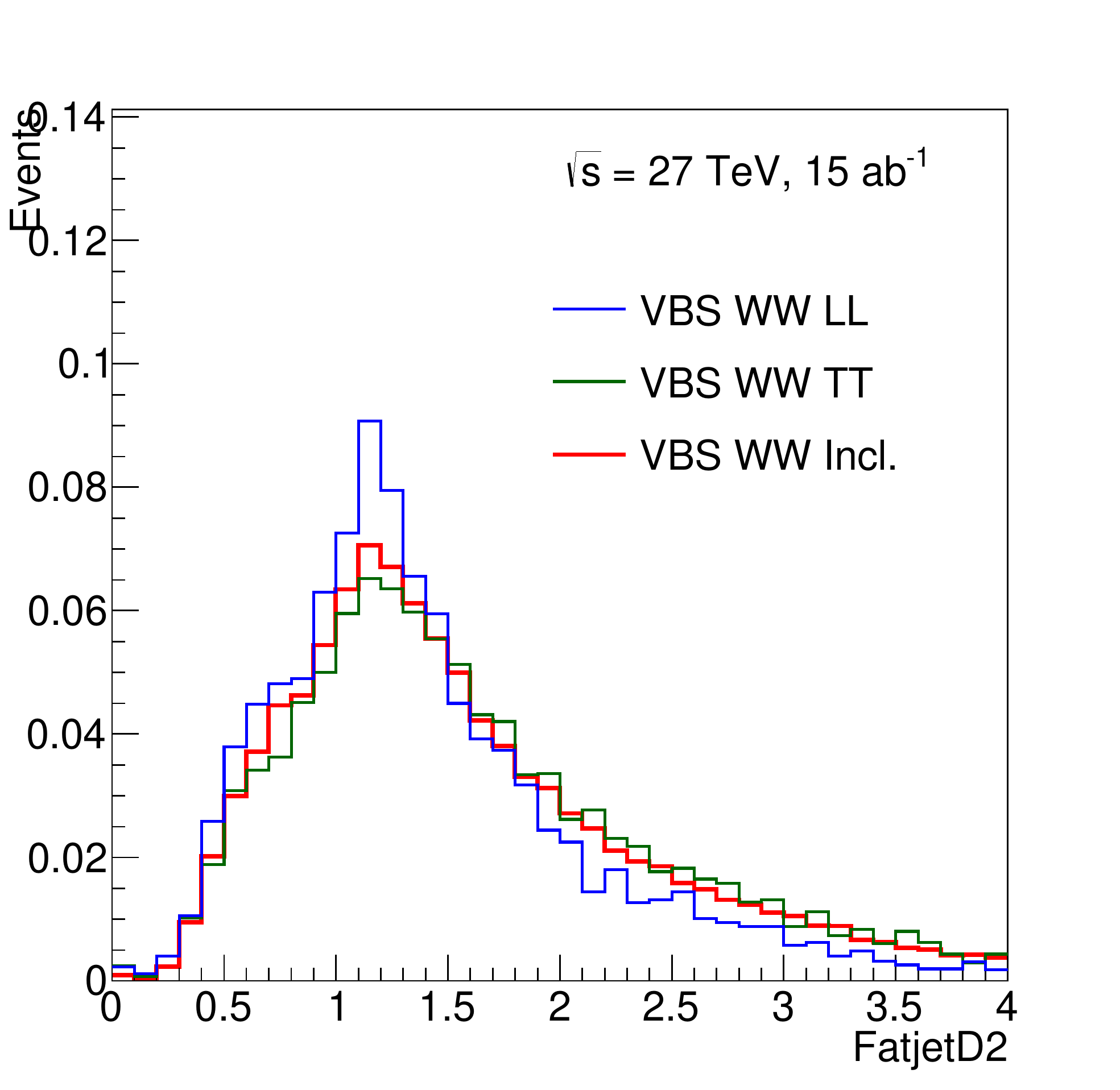}
\includegraphics[width=0.4\textwidth]{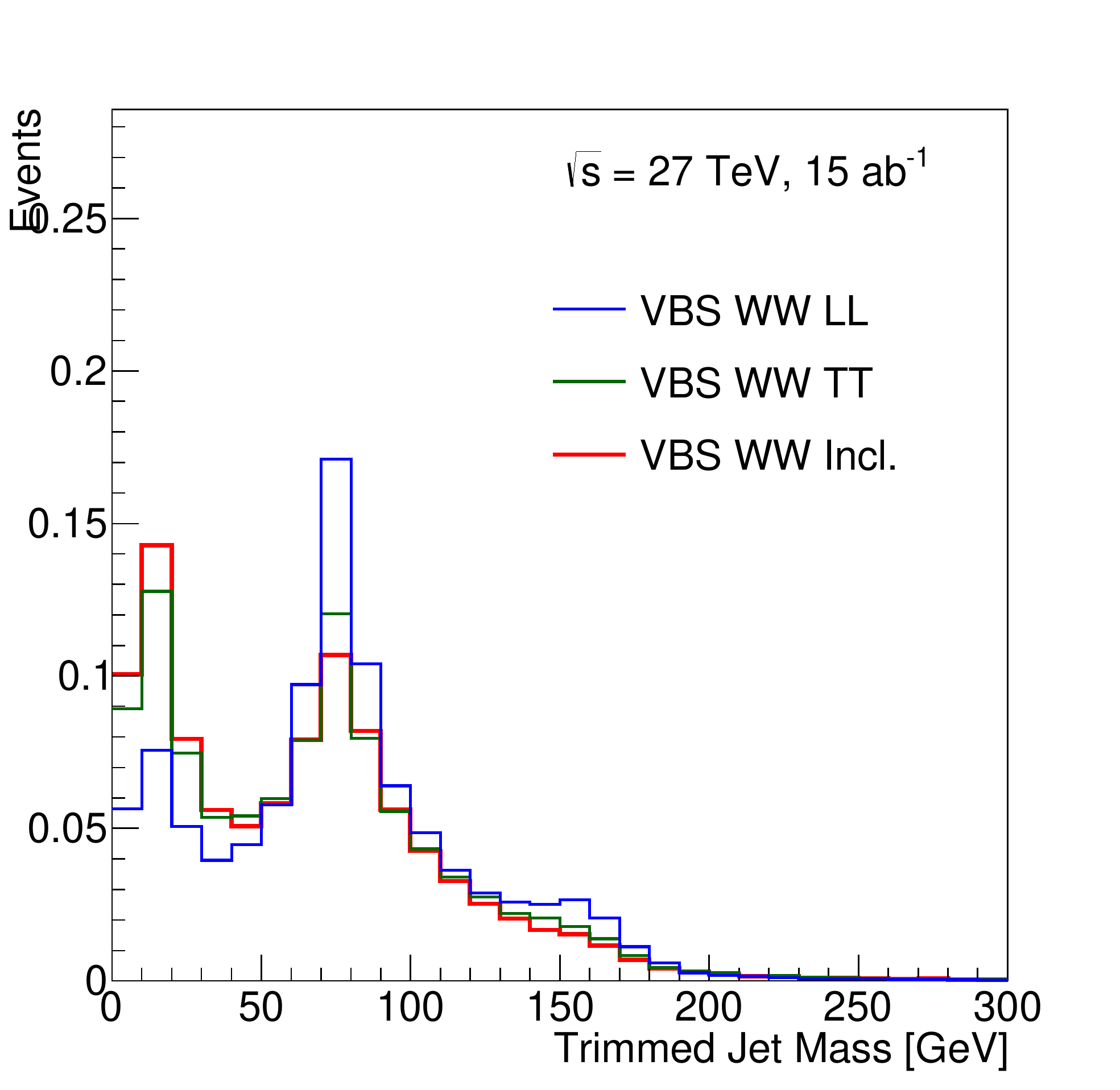}\\
\includegraphics[width=0.4\textwidth]{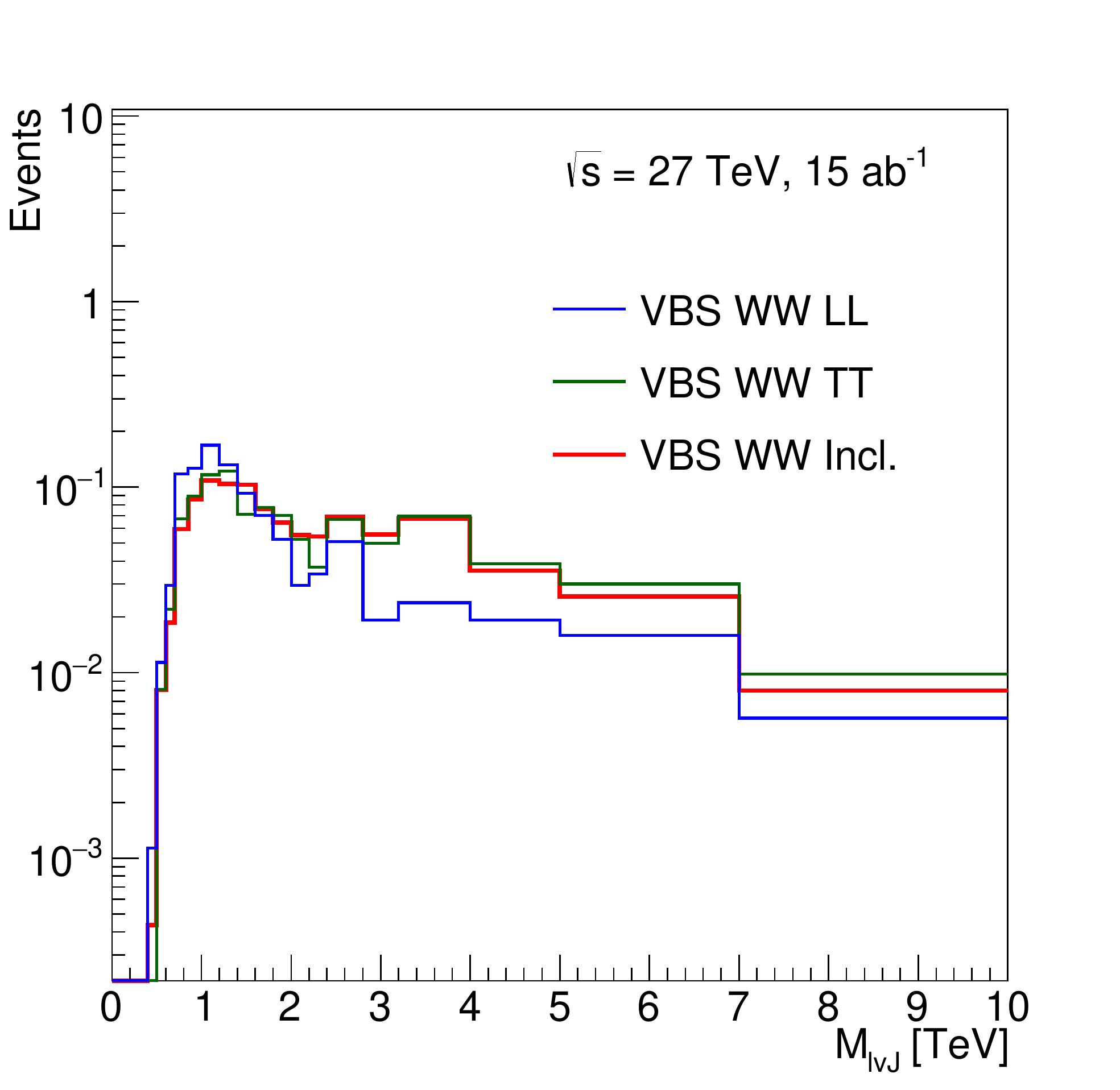}
\includegraphics[width=0.4\textwidth]{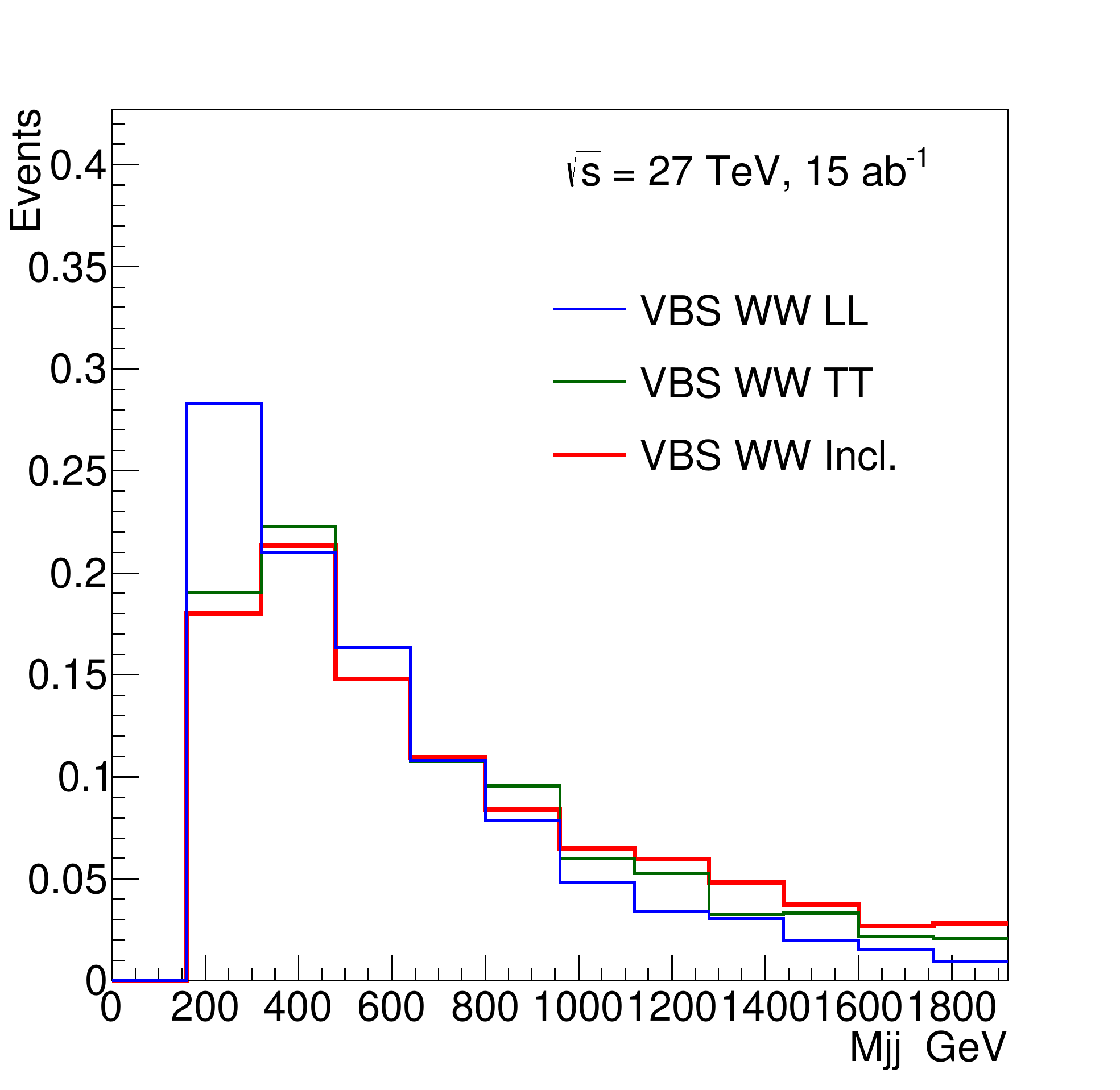}
\caption{Normalized distributions of $D_2$, trimmed large-$R$ jet mass, the mass of the VBS system $m_{WWjj}$ and the mass of the tagging jet system for the longitudinally-polarized (LL) and transversely-polarized (TT) VBS $WW$ events as well as the inclusive VBS $WW$ events. Only the distributions for the merged analysis are shown. 
}
\label{fig:KinematicVBSLongitudinal1}
\end{figure}

The VBS events are expected to be produced with three different polarization states of 
the final-state vector bosons: both of them being longitudinally (LL) or transversely polarized (TT), or in the mixed polarization
state (LT). 
To determine the expected sensitivity to the longitudinally-polarized electroweak 
$WW \rightarrow lvqq$ scattering (denoted by $W_LW_L$), a large
set of kinematic variables has been investigated to provide the best separation from 
the TT or LT component. The normalized distributions of the selected variables used for the merged analysis are shown in Fig.~\ref{fig:KinematicVBSLongitudinal1}, illustrating the differences between the LL or TT component from the inclusive VBS events.
The single most powerful discriminating variable is the total invariant mass of the system 
$m_{WWjj}$
composed of the diboson and the two tagging jets.
A BDT is built using the variables shown in Fig.~\ref{fig:KinematicVBSLongitudinal1} to take full advantage of the kinematic differences. The BDT is trained such that the LL component is discriminated as a signal from the LT and TT components as well as the rest of the background sources.%
The discriminating power for the LL component is potentially improved if the different background components are treated separately with dedicated multivariate techniques.

\begin{figure}
\centering
\includegraphics[width=0.45\textwidth]{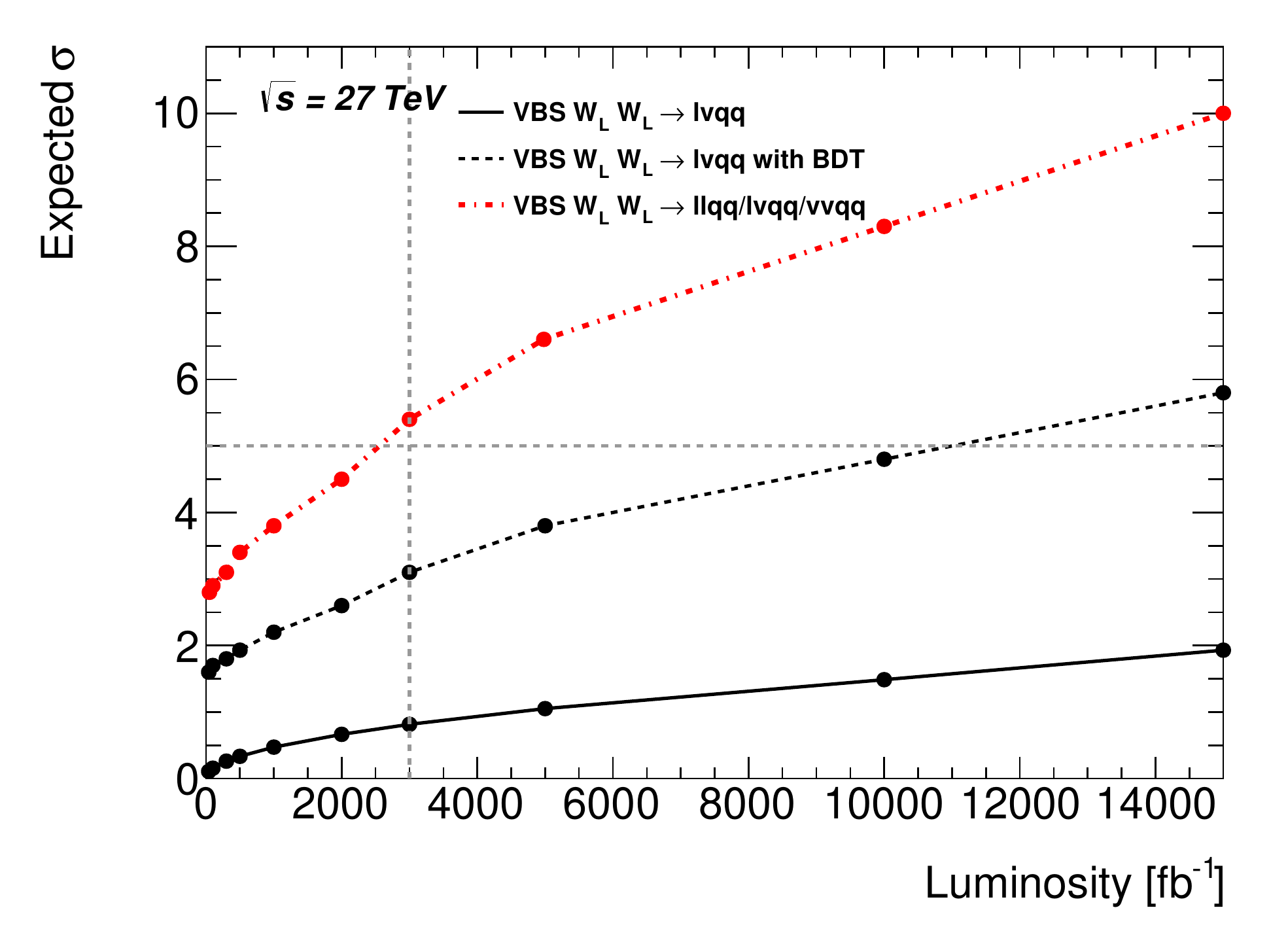}
\includegraphics[width=0.45\textwidth]{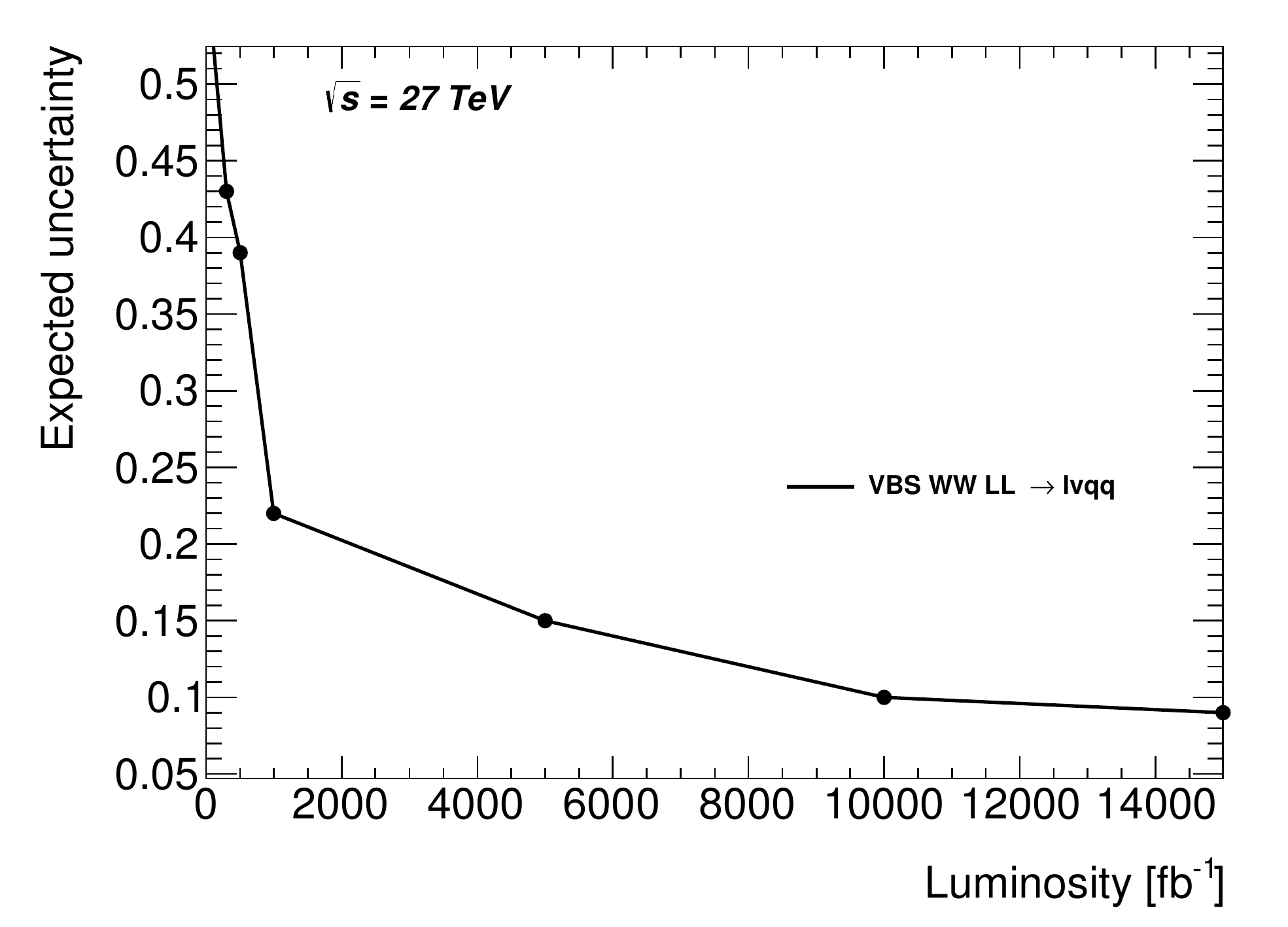}
\caption{Observed significance as a function of integrated luminosity (left) and expected cross-section uncertainty (right) for the VBS $W_LW_L$ signal, assuming a 10\% $W_LW_L$ fraction predicted by the MadGraph generator, in the $\ell\nu qq$ channel at $\sqrt{s}=27$~TeV. The solid and dashed lines on the left shows the expected significance obtained by fitting to the total invariant mass of the VBS system and the BDT output, respectively. The dot-dashed line shows the expected significance from the combination of all the three semi-leptonic channels assumed to have sensitivity similar to the $\ell\nu qq$ channel. 
}
\label{fig:VBSLL}
\end{figure}

The expected significance of the extracted electroweak $W_LW_L\rightarrow lvqq$ signal is 
extracted by performing a simultaneous binned maximum-likelihood fit to the BDT distributions 
in the signal regions and the $W$+jets and $t \bar{t}$ control regions. 
A test statistic based on the profile likelihood ratio is used to test hypothesized values of the longitudinal scattering cross-sections. 
The likelihood is defined as the product of the Poisson likelihoods for all signal and control regions for a given analysis channel. 
The fit includes four background contributions from $W$+jets, $t \bar{t}$, QCD diboson and the electroweak LT and TT $WW$ processes. The treatment of systematic uncertainties is similar to that used in the VBS significance estimate above. 
The result of the expected significance is shown in Fig.~\ref{fig:VBSLL}. Also shown is the significance
obtained by fitting only to the $m_{WWjj}$ variable. The figure shows that the BDT will bring a significant improvement in the analysis sensitivity, resulting in the expected significance of about 3 (5)$\sigma$ with
an integrated luminosity of 3 (12)~ab$^{-1}$. 
In addition to the $\ell\nu qq$ channel, the expected combined sensitivity with the other semi-leptonic channels, $\ell\ell qq$ and $\nu\nu qq$, is shown as well, assuming an equal sensitivity for the individual channel to the $\ell\nu qq$.
Figure~\ref{fig:VBSLL} also shows the expected cross-section uncertainty of the electroweak $W_LW_L$ scattering 
as a function of integrated luminosity. 

\section{Conclusion}
The prospects of searches for new heavy resonances decaying to diboson ($WW$) and measurements of electroweak $WW/WZ$ production via vector boson scattering (VBS)  in association with a high-mass 
dijet system in the $\ell\nu qq$  final states are presented.
The results are based on an integrated luminosity of 15~ab$^{-1}$ of proton-proton collisions at 
$\sqrt{s}=27$~TeV with an ATLAS-like detector simulated in the Delphes framework. 
The cross-section measurement of the electroweak $WW/WZ$ production in VBS processes 
is expected to reach the precision of $\sim$2-3\%, improving the expected accuracy at the HL-LHC by a factor of 2. 
Probing the longitudinal component of the electroweak $WW/WZ$ production is of paramount importance 
at future colliders.With 3~ab$^{-1}$ of data at $\sqrt{s}=27$~TeV, the HE-LHC could separate the longitudinal component
with the significance of $\sim3\sigma$ for the single $\ell\nu qq$ channel and  $\sim5\sigma$ for all the semi-leptonic channels combined.
The diboson resonance searches are interpreted for sensitivity to a simplified phenomenological model 
with a heavy gauge boson. The discovery reach at the HE-LHC is extended to 8~TeV with 15~ab$^{-1}$ 
of data at $\sqrt{s}=27$~TeV, more than two-fold increase from the expectation at the HL-LHC with 3~ab$^{-1}$ at $\sqrt{s}=14$~TeV.

\clearpage

\bibliographystyle{ieeetr}

\end{document}